\documentclass{article}

\PassOptionsToPackage{hyphens}{url}
\usepackage{url}
\usepackage[T2A]{fontenc} %
\usepackage[english]{babel} %
\usepackage[auth-lg, max2]{authblk}

\usepackage[letterpaper,top=2cm,bottom=2cm,left=3cm,right=3cm,marginparwidth=1.75cm]{geometry}

\usepackage{amsmath}
\usepackage{graphicx}
\usepackage[colorlinks=true, allcolors=blue]{hyperref}

\title{Sovereign Large Language Models: Advantages, Strategy and Regulations}

\author[1]{%
Mykhailo Bondarenko\thanks{mykhailo.bondarenko@ucu.edu.ua}%
}
\author[1]{%
Sviatoslav Lushnei\thanks{sviatoslav.lushnei@ucu.edu.ua}%
}
\author[1]{%
Yurii Paniv\thanks{paniv@ucu.edu.ua}%
}
\author[1]{%
Oleksii Molchanovsky\thanks{olexiim@ucu.edu.ua}%
}
\author[3]{%
Mariana Romanyshyn\thanks{mariana.scorp@gmail.com}%
}
\author[2]{%
Yurii Filipchuk\thanks{yura.filipchuk@gmail.com}%
}
\author[2]{%
Artur Kiulian\thanks{akiulian@gmail.com}%
}

\affil[1]{Ukrainian Catholic University}
\affil[2]{OpenBabylon Inc, openbabylon.com}
\affil[3]{lang-uk, lang.org.ua}

\date{January 15, 2025}

\begin{document}

\maketitle

\section{Executive Summary}

This report analyzes key trends, challenges, risks, and opportunities associated with the development of Large Language Models (LLMs) globally. It examines national experiences in developing LLMs and assesses the feasibility of investment in this sector. Additionally, the report explores strategies for implementing, regulating, and financing AI projects at the state level.

International experiences indicate that LLMs significantly enhance administrative efficiency. In regulatory processes, they streamline the management of legal documents (Albania, Serbia), facilitate communication between government authorities and citizens (Netherlands), and support public procurement and legal translations (Albania). In social services, LLMs assist with agricultural advisory services (Nigeria) and provide targeted information on state aid programs (North Macedonia). In healthcare, they contribute to structuring medical records, predicting risks, and generating recommendations (Sweden, Kenya). In defense and security, LLMs enhance open-source intelligence (OSINT) analysis and generate responses adapted to military communication protocols (USA, China).

By 2024, government adoption of AI had increased by over 1.5 times compared to 2021, driven by the growing relevance of national LLMs, which outperform global models on localized benchmarks (Brazil, Albania). National models provide enhanced data protection, safeguard national security interests, and contribute to maintaining international competitiveness.

The first national AI strategies, introduced between 2019 and 2020, are undergoing revision, particularly concerning LLM development. Key strategic priorities include:
\begin{enumerate}
    \item Strategic autonomy: Reducing reliance on foreign technologies, expanding domestic research capacity, and fostering innovation.
    \item Ethical and legal compliance: Aligning with international AI principles, ensuring data protection, and establishing public oversight mechanisms.
    \item Infrastructure development: Establishing supercomputing facilities (India, Brazil, Israel), supporting data centers, and expanding access to high-performance computing (EuroHPC program).
    \item Data accessibility: Promoting open data platforms (Japan), regulating data exchange and usage (Saudi Arabia), and funding data initiatives through grants (Horizon Europe, CLARIN).
    \item Talent and innovation investment: Supporting AI startups, funding universities and research institutions, and fostering workforce development (Thailand, Brazil, Saudi Arabia).
\end{enumerate}

Investment in AI development at the national level includes significant financial commitments. The European Union allocates €1 billion annually through 2027, while the Netherlands has invested €204.5 million. Brazil has committed \$4.2 billion, with \$200 million dedicated to the development of a national LLM.

Funding sources for national LLMs vary, including government allocations (Brazil, Sweden, Singapore, Thailand), corporate investments (Samsung, Baidu), private sector initiatives (Mistral in France), and industry-backed projects (Saudi Aramco in Saudi Arabia). Grant programs, such as EU CLARIN and Horizon Europe, also contribute. Hybrid funding models combining state, corporate, and private investments are common (Thailand, Greece).

LLM development is carried out through collaborations among research institutions, universities, and private-sector partners (Japan, Saudi Arabia). Models are either developed from scratch or fine-tuned using existing architectures (e.g., LLaMA). Dedicated working groups and oversight bodies ensure project coordination and ethical compliance (Netherlands).

Regulatory frameworks for AI oversight include auditing AI applications (Brazil), content moderation mechanisms (China), licensing AI solutions for industry-specific use (South Korea), and establishing regulatory bodies for policy development and data protection (Turkey, Saudi Arabia, Netherlands).

Coordinated investments, infrastructure development, and regulatory frameworks foster the development of national LLMs, enabling the adaptation of AI solutions to local contexts while ensuring data security and preserving cultural and linguistic diversity.

\section{Introduction}

Artificial intelligence (AI) has become one of the most influential technologies of the 21st century, transforming multiple industries such as education, electronic governance (e-governance), and medicine. The term AI encompasses numerous subfields, such as computer vision, natural language processing (NLP), generative AI, and others. In recent years, large language models (LLMs), a subset of generative AI capable of analyzing text and generating new content, have gathered significant public attention. The development of such models was made possible by the rapid increase of available data and computing power. These factors have led to the emergence of big data and the creation of specialized, high-performance supercomputing infrastructures utilizing devices such as graphics processing units (GPUs) and tensor processing units (TPUs).

A pivotal moment in the evolution of generative AI was the launch of ChatGPT \cite{openai_chatgpt}, which, for the first time, made generative language models accessible to the general public. Since the company has decided to keep its model private, the largest companies and teams around the world tried to replicate their success by training their own models. However, their development was hindered by the need for vast computational power and data resources. To address this, countries worldwide have begun investing in infrastructure for computational and data centers, forming the backbone for the development of modern LLMs. Many companies and researchers with limited resources also utilized fine-tuning methods, reducing the cost of training by building upon pre-existing open-source models.

Nevertheless, as LLMs evolve rapidly, new challenges arise. These include growing concerns about the safe use of AI and the monopolization of its development, prompting nations to develop their national strategies. These strategies emphasize the principles of responsible AI, which focuses on the ethical development of models, and include investments in projects to create language models tailored to local contexts, cultures, and applications.

In this report, we analyze the key trends, challenges, risks, and opportunities associated with the development of large language models. We will explore the various applications of such models, including their role in national security, as well as the potential for international cooperation, including programs available to Ukraine. Additionally, we will examine strategies for the implementation of AI technologies, the role of investments, and the associated challenges.

To ensure relevance to Ukraine, we will focus beyond the key players in LLM space, such as the United States and China. By doing so, we aim to highlight examples that may better align with Ukraine's context and experience, as opposed to those of nations already holding leading global roles.

The structure of this report is as follows: the second chapter will discuss the applications of language models, particularly in national security, and their economic impact. The third chapter will focus on trends shaping national AI strategies, the LLM development process, and related examples. The fourth chapter will concentrate on regulations, with special attention to emerging trends. Finally, the fifth chapter will address international cooperation, including EU programs (Ukraine is an accession candidate to the EU). The sixth chapter will provide conclusions based on the findings of this work.

\section{Methodology}
Considering the global technological leadership of the USA, the authors analyzed which strategies different countries use to reduce the gap, examining the regulatory environment, business approaches to implementation, approaches to government funding, and so on. The authors primarily looked at high and middle-income countries, their independent approaches to creating sovereign large language models and AI development in general, and international cooperation programs between them. More than 18 macroregions were analyzed, covering about 85\% of the world's population. For each country, the authors collected answers to the following questions:
\begin{itemize}
\item Who trains large language models in the country?
\item How is their training and use regulated?
\item How is the use of training data regulated?
\item How and by whom is the training funded?
\item How are universities, private companies, and the government involved?
\item Which international programs have they joined?
\end{itemize}
The authors subsequently grouped and systematized the answers to these questions into common approaches and themes in the report.
\newpage
\tableofcontents
\newpage

\section{Why Sovereign LLMs?}

In this section, we will explore the prospects associated with developing a sovereign language model. Specifically, we will examine the applications of language models in subsection \ref{subs:usages}, the impact of developing a sovereign language model on national security in subsection \ref{subs:defence_usage}, and the economic implications in subsection \ref{subs:economics_usage}. In both sections, we will focus on projects supported by the respective governments. This section aims to provide as many examples as possible to highlight the main opportunities and challenges related to building a sovereign language model.

\subsection{Use Cases For Large Language Models} \label{subs:usages}

Language model-based artificial intelligence is already being applied in various areas of life, such as education and e-governance. This section presents such initiatives, some of which could serve as a reference for Ukraine. The applications discussed in this section highlight where LLMs are most beneficial, demonstrate usage trends, and shed light on their limitations.

\begin{itemize}

    \item \textbf{Education.} The initiatives discussed below focus on integrating language models into the educational process to enhance efficiency. Although this field is still in its early stages, and these technologies are just beginning to be implemented in practice, models like BgGPT demonstrate significant educational potential.

    For example, one of the objectives of BgGPT, an LLM trained on data that consisted of predominantly Bulgarian text, is to shift the perception of artificial intelligence in educational institutions and integrate it into learning processes, as emphasized by Professor Galin Tsokov, Minister of Education and Science of Bulgaria \cite{atanasova_2024_bulgaria}. BgGPT can perform tasks and answer questions related to the learning process, similar to OpenAI’s ChatGPT and Google’s Bard \cite{atanasova_2024_bggpt}, potentially reducing the workload for teachers. In a study conducted in December 2024, BgGPT demonstrated some of the best results when tested on school exams provided by the Ministry of Education \cite{alexandrov_2024_bggpt}. In some cases, its performance surpassed popular language models like OpenAI’s ChatGPT and Meta’s LLama, especially compared to models of the same size.

    In Greece, the Meltemi language model was developed \cite{a2024_meltemi}, with one of its goals being to serve as a useful tool in educational settings. To achieve this, the digital assistant can interact with students, solve problems from their learning materials, create exercises specific to their needs, explain terms, and even simplify specific textbook texts \cite{a2024_meltemi}.

    India’s National AI Strategy, published in 2018, highlighted education as one of the five sectors projected to benefit the most from the advancement of artificial intelligence \cite{roy_2018_national}.
    
    The IndiaAI Mission, which received an equivalent of \$1.2 billion in government investment in 2024 \cite{a2024_cabinet}, developed BharatGen \cite{a2024_bharatgen}, an Indian LLM aimed at making AI technologies accessible to foster innovation in key sectors, including education \cite{bharatgen_2024_about}.
    
    Additionally, a private project called The Indus Project by Tech Mahindra, in collaboration with NVIDIA, is developing an open-source LLM for Indian languages, and lists education aid for students as one of the potential benefits \cite{a2024_the}.

    In Singapore, the AI Centre for Educational Technologies (AICET) \cite{a2024_home} collaborates with the Ministry of Education to implement projects to enhance the educational system. This center was established by AI Singapore, which developed the Singaporean language model SEA-LION \cite{aisingapore_2023_sealion}, and funded by Smart Nation Singapore \cite{smartnationsingapore_the}, an initiative of the Ministry of Digital Development and Information (MDDI). Among AICET’s projects is the Codaveri automated feedback system \cite{codaveri}, a language model trained to assist students in learning programming. The system provides comments with hints to guide students toward the correct answer rather than directly giving the solution \cite{liew_2023_3}, making the learning process more effective.

    Also in Singapore, under a five-year plan titled Artificial Intelligence at the National Institute of Education (AI@NIE) \cite{a2022_introducing}, the National Institute of Education will invest in research and innovation to harness artificial intelligence in education \cite{a2023_shockwaves}. The initiative aims to prepare Singaporean educators for the effective use of AI in teaching and to promote research into innovative and ethical applications of AI in education.

    Finally, in collaboration with Google, Nanyang Polytechnic has developed Course AutoBot, a tool built with LLMs to aid teachers by helping them create course material \cite{accesspartnership_2024_strengthening}.

    \item \textbf{Electronic governance.} The initiatives listed below aim to simplify government services and improve their accessibility by using large language models and artificial intelligence in general.

    One of the leading European countries in the use of LLMs in e-governance is Albania. As part of its digitalization efforts, Albania launched Virtual Assistant 1.0 in December 2023, followed by Virtual Assistant 2.0 in 2024 \cite{a2025_aibased}. The latest version of this assistant offers functionality to automate access to certain documents and public services by fully covering some steps of the application submission process. To achieve this goal, this chatbot will be integrated into the e-Albania digital governance portal.

    Prime Minister Edi Rama has stated that in the future, artificial intelligence will replace a significant portion of services at local and national government levels, particularly in public procurement, to combat corruption \cite{taylor_2024_albania}.

    Additionally, the National Agency for Information Society (AKSHI) announced a tender with a financial limit of approximately €2.6 million titled ''Use of Artificial Intelligence in the Process of Transposing the Acquis for European Integration'' \cite{a2023_tender}. As a result of this initiative, Albania plans to develop a language model to accelerate EU accession by translating the required legal documents.

    With a similar motivation, COMtext.SR was developed in Serbia, a text model built to automatically analyze legal documents \cite{a2018_comtextsr}. Created at the Innovation Center of the Faculty of Electrical Engineering at the University of Belgrade, this project focuses on legal texts, a domain not yet covered by existing tools for the Serbian language. Analyzing legal documents is important for public administration, NGOs, and companies, especially for EU integration and alignment with EU standards \cite{kovacevic_2024_university}.

    To implement this project, the center was equipped with infrastructure partially funded by a €200 million loan from the European Investment Bank. The project also received support from the EU Instrument for Pre-accession Assistance, the Council of Europe Development Bank, and the Serbian government \cite{kovacevic_2024_university}.

    In North Macedonia, as part of its digitalization efforts, the ADA assistant was created \cite{a2023_bytyqi}. ADA is an AI-based digital assistant based on LLMs, capable of answering the most frequently asked questions about government assistance programs provided by the Republic’s Government \cite{a2023_ada}.

    In the Netherlands, the Netherlands Organisation for Applied Scientific Research (TNO) \cite{TNO2023} has studied the use of artificial intelligence in the public sector \cite{TNO2023_press}. According to their research, as of 2024, the use of AI applications by the government increased by more than 1.5 times compared to 2021. Of the 266 identified AI applications, 105 (39\%) were utilized by municipalities \cite{Hoekstra2025Quickscan}.

    Furthermore, 49 technologies were identified for automatic speech, text recognition, and natural language processing (NLP). These applications were primarily used for anonymizing personal and confidential financial data. The development of anonymization software is significantly influenced by the Open Government Act (Wet open Overheid, WOO), which requires the disclosure of various official documents, some of which need to be anonymized before publishing \cite{gemeentebreda_open}.

    Additionally, a virtual assistant named Codi, built on generative language models, is being developed to answer so-called Parliamentary Questions \cite{nlaicoalition_2024_codi}. This mechanism, which is key to the Netherlands democracy, allows citizens to submit questions to Parliament, to which it is obliged to respond. Each year, Parliament receives more than 3,000 questions, requiring thousands of civil servants to draft responses. To optimize this process, Codi will gather information from existing documents and previous questions to provide the necessary data to help public officials do this work.

    \item \textbf{The aging population problem.} In Sweden, generative artificial intelligence is being proposed to address the challenges of an aging population. As the population ages, one of the significant consequences is the reduction of the available workforce in many parts of the public sector, especially where the work involves a lot of text processing and report generation. This challenge was one of the challenges that motivated the development of the GPT-SW3 language model, designed to be integrated into the public sector to use the power of LLMs for handling such text-based tasks \cite{a2024_a}.

    \item \textbf{Farming and agriculture.} The initiatives listed below aim to leverage language models to assist farmers in agriculture.

    In India, the developers of the BharatGen language model, which was discussed above and received significant government investment, see innovation in the agricultural sector as one of their primary goals \cite{a2024_bharatgen}. The development of this model aligns with India’s 2018 National AI Strategy, which identified improving agricultural productivity and increasing farmers' incomes as key areas for AI application in the agricultural sector \cite{roy_2018_national}.

    In Nigeria, where agriculture is considered a primary focus for AI applications \cite{NAIS2024}, the company Crop2Cash has developed a voice-based chatbot called FarmAdvise. This chatbot, based on a combination of LLMs, voice generation, and recognition, operates as a free hotline for smallholder farmers, enabling them to receive valuable knowledge, improve farming skills, make informed decisions about crop cultivation and livestock management, and enhance productivity and profitability of their farms \cite{crop2cash_our}. This initiative is particularly relevant in Nigeria, where many citizens lack stable internet access \cite{nigeriancommunicationscommission_2025_industry}.

    As a final example, in a somewhat unconventional usage, the company Instadeep from Tunisia developed AgroNT, an LLM trained on the genomes of crops to automate their study, assisting researchers in the field of crop improvement and agricultural research \cite{Mendoza-Revilla2024}.

    \item \textbf{Medicine.} The initiatives listed below aim to apply language models in the healthcare sector.

    In Sweden, the Halland Hospital Group tests the GPT-SW3 language model for medical applications. In this region, the model is being used for tasks such as summarizing medical records, predicting adverse events, coding journals, extracting information from unstructured text, and creating patient discharge messages \cite{a2023_open}. Markus Lingman, senior physician, professor, and strategic director at the Halland Hospital Group, notes that this generative language model opens up new possibilities and emphasizes the value of having complete control over the model itself. He believes language models can significantly assist the healthcare field \cite{a2023_open}.

    In Kenya, the language model UlizaLlama (AskLlama) \cite{jacarandahealth_2024_jacaranda}, developed by Jacaranda Health \cite{jacarandahealth_2023_about}, is designed to provide medical advice, with a focus on supporting young and expecting mothers across Africa. Due to the sensitive nature of such advice, the company employs human reviewers to verify and edit responses before they are sent to mothers. Nonetheless, the language model significantly simplifies this process. Notably, the company emphasizes that teams deploying their application maintain full control over user data, which is crucial for ensuring medical confidentiality, which is being trusted to unverified vendors when using off-the-shelf models like ChatGPT \cite{jacarandahealth_2024_jacaranda}.
\end{itemize}

\subsection{Language Models for National Security} \label{subs:defence_usage}

Developing a sovereign large language model can play a supportive role in national security. Specifically, subsection \ref{subsubs:nd_data_safety} will examine examples of countries that have integrated artificial intelligence, including language models, into their government services, highlighting their motivation to protect citizens' data. Additionally, subsection \ref{subsubs:nd_defence} will explore existing applications in the defense sector.

\subsubsection{Data Security} \label{subsubs:nd_data_safety}

\begin{itemize}

    \item \textbf{Belgium.} Professor Martin Vechev, the founder of the Bulgarian institute INSAIT \cite{a2024_governance}, emphasized at a press conference \cite{a2025_bulgaria} that BgGPT, the Bulgarian language model that received a \$100M investment from the government, ensures data storage and processing within the country. This is a strategic advantage for maintaining data security and technological sovereignty.

    \item \textbf{Albania.} In December 2023, Albania launched Virtual Assistant 1.0, followed by Virtual Assistant 2.0 in 2024 \cite{a2025_aibased}, designed to assist citizens and users of the e-Albania portal. The latest version of this assistant automates the application process for obtaining specific documents and government services. Sending documents and personal data to uncertified services poses risks to information security, making a sovereign language model critical for such applications.

    \item \textbf{Germany.} In Germany, the startup Aleph Alpha \cite{a2025_aleph}, one of the leaders in AI development in Europe, launched a project called PhariaAI in 2024. This project enables public institutions and businesses to integrate artificial intelligence, including language models, into their processes. The startup emphasizes that the key advantage of its solution lies in the control over sensitive data, which is partly ensured by using proprietary language models within a controlled environment.

    \item \textbf{Other usages.} Data security is critically important not only to government institutions. For instance, the medical applications mentioned in subsection \ref{subs:usages}, which require compliance with medical confidentiality, cannot operate on uncertified servers. Therefore, when using LLMs, the control over the underlying models is crucial.
\end{itemize}

\subsubsection{National Defence} \label{subsubs:nd_defence}

Language models have also found applications in the defense sector. A declassified NATO document dated January 9, 2024, determines activities related to assessing the risks and opportunities associated with large language models \cite{natoscienceandtechnologyorganization_2024_stoactivities}. Task Force Lima, established in 2023 in the United States to identify applications of generative AI in defense, identified over 180 use cases where such tools could be beneficial \cite{vincent_2023_inside}. Following the completion of this research, the Artificial Intelligence Rapid Capabilities Cell (AI RCC) was launched to develop some of these applications \cite{harper_2024_pentagon}. According to Radha Plumb, head of the Chief Digital and AI Office (CDAO), these applications include command and control, decision support, and enterprise management functions such as financial and healthcare information management \cite{harper_2024_pentagon}. In this subsection, we will explore a subset of these applications that are publicly known.

\begin{itemize}

    \item \textbf{Defense LLAMA.} Defense LLAMA is a project by the U.S.-based startup ScaleAI, which, building on Meta's open LLAMA model \cite{meta_llama}, trained a language model capable of answering defense-related questions. The Defense LLAMA model was trained on a vast dataset that includes military doctrine, international humanitarian law, and policies aligned with the U.S. Department of Defense (DoD) guidelines on armed conflicts, as well as the DoD’s Ethical Principles for Artificial Intelligence. Defense LLAMA responds consistently with the style and tone of the Office of the Director of National Intelligence (ODNI), enhancing its relevance to its users. The developers emphasize that Defense LLAMA is available exclusively in a U.S. government-controlled environment as part of Scale AI’s project named Scale Donovan \cite{scaleai_introducing}.

    \item \textbf{USAF.} According to research conducted by the Air Force Institute of Technology, the United States Air Force (USAF) is already employing LLMs for wargaming and automated summarization tasks, illustrating their potential to optimize operations and support decision-making processes \cite{caballero_2024_on}.

    \item \textbf{ChatBIT.} The Chinese model ChatBIT, similar to Defense LLAMA, is based on one of Meta's LLAMA models \cite{meta_llama} and has been adapted for military purposes. Specifically, this model has been utilized for open-source intelligence (OSINT) and tasks aimed at simplifying military communication \cite{cheung_2024_prc}. Given that China recently developed its language model, Qwen \cite{qwenteam_2024_about}, future versions of ChatBIT may be built on top of it.

\end{itemize}

\subsection{How Will AI Integration Impact the Economy?} \label{subs:economics_usage}

In 2023 and 2024, artificial intelligence emerged as one of the key drivers of economic growth worldwide, with the stocks of companies developing AI solutions experiencing the fastest growth \cite{laidley_2025_what}. In this section, we will share statistics from various studies illustrating AI's future potential economic impact. A crucial point emphasized across these studies is that to achieve the best possible results, the integration of AI into the economy and business practices must be done correctly. In addition to providing economic indicators, we will share some of the recommendations highlighted in these studies.

\begin{itemize}

    \item \textbf{Global impact.} According to a 2023 study by McKinsey \& Company, generative AI has the potential to contribute between \$2.6 trillion and \$4.4 trillion annually to the global economy across 63 applications analyzed by the firm \cite{mckinseycompany_2023_economic}.

    Approximately 75\% of this estimated value is attributed to four key areas of generative AI applications: customer operations, marketing and sales, software engineering, and research and development (R\&D) \cite{mckinseycompany_2023_economic}.

    The study also projects that generative AI alone could boost labor productivity by 0.1–0.6 percent annually through 2040 by reallocating workers' time to other activities. When combined with other technologies, productivity growth could range from 0.5 to 3.4 percentage points. However, achieving these gains will require workers to adapt to new skills, and some may need to transition to different professions, posing a significant risk \cite{mckinseycompany_2023_economic}.

    item \textbf{Asia-Pacific Region.} According to a study conducted by Access Partnership on behalf of Google, local businesses in the Asia-Pacific region could add \$3 trillion to the economy by successfully integrating AI-based solutions, including generative AI \cite{a2024_strengthening}. This is the combined estimate for 14 key Asia-Pacific countries: Australia, Hong Kong, India, Indonesia, Japan, Malaysia, New Zealand, Pakistan, the Philippines, Singapore, South Korea, Taiwan, Thailand, and Vietnam.

    Singapore is expected to become central to this development, with local businesses projected to generate an additional \$147.6 billion by 2030. In a survey conducted as part of the study, 78\% of Singaporeans expressed interest in learning more about AI, and more than half already use generative AI tools. Furthermore, the study found that 84\% of individuals aged 18 to 35 use AI tools (including generative AI), and 94\% of surveyed employees expect to utilize generative AI tools in their workplace within the next five years \cite{a2024_strengthening}.

    The study highlights that one of the key components driving AI adoption in Singapore is the availability of skilled AI professionals \cite{a2024_strengthening}. It also notes that Singapore plans to increase the workforce by around 15,000 such professionals by 2028, citing the country’s 2024 National AI Strategy \cite{internationaltradeadministration_2024_singapore, a2024_strengthening}.

    \item \textbf{Europe.} According to a study by Strand Partners, which was commissioned by Amazon Web Services (AWS), the integration of AI by European businesses is expected to contribute €600 billion to the economy by 2030 \cite{a2024_ai_2, brown_2024_ai}. However, the report highlights three key challenges that European countries must address to fully unlock AI's potential: fostering an innovation-friendly environment, overcoming Europe's digital skills gap, and ensuring that businesses of all sizes can access the latest technologies.

    \item \textbf{Thailand.} According to a survey by SCB X-SCB EIC on the readiness of small and medium-sized enterprises (SMEs) to adopt generative AI in Thailand, conducted from April 18 to 25, 2024, it is estimated that generative AI will boost Thailand's GDP by 6\% by 2030 \cite{a2024_genai, a2024_outlook}.

    The survey notes that Thai SMEs are not fully prepared for the arrival of GenAI, particularly regarding their data and infrastructure readiness. The survey highlights challenges faced by early adopters, such as a lack of skilled personnel (41\%), cost management related to AI development and deployment (30\%), and the absence of clear criteria for evaluating potential AI solutions (29\%).

    \item \textbf{India.} Another study, also conducted by Access Partnership on behalf of Google, estimates AI's impact on India's economy at an equivalent of \$400 billion by 2030 \cite{a2023_economic}. In particular, the study highlights how the Indian Institute of Science (IISc) collaborated with Google India under the Vaani project to collect language data across 773 districts nationwide to aid the development of an LLM for India. The report also underscores how Google's Cloud Security AI Workbench, part of which relies on LLMs, enhances cybersecurity in the country.

\end{itemize}

\section{How to develop sovereign LLM?}

In this section, we will examine the development of sovereign language models from various perspectives. Specifically, we will review and analyze national strategies as a critical tool for declaring the objectives of AI development and creating language models in subsection \ref{subsec:NationalStrategy}. Additionally, we will explore the details and patterns involved in the creation of language models across different initiatives in subsection \ref{subsec:ModelDevelopment}, as well as the methods and scale of funding allocation for AI development, particularly for large language models (LLMs), along with other indirect investments in subsection \ref{subsec:ModelInvestments}.

\subsection{National Strategies}
\label{subsec:NationalStrategy}

National AI strategies are a cornerstone of advancing sovereign artificial intelligence, especially in developing large language models (LLMs). These strategies act as a guiding framework for governments, setting priorities, allocating resources, and charting the course for technological progress in AI. Beyond driving research and innovation, they establish a clear vision, define objectives, and outline the requirements necessary to effectively integrate AI solutions into various aspects of everyday life.

Most governments introduced national AI strategies or plans during 2019–2020. The primary goals and motivations behind these plans can be categorized into several key areas, although some countries may adopt a combination of the approaches described below:

\begin{itemize}
\item \textbf{Artificial Intelligence as a Strategic Industry.} Many nations recognize AI as a vital sector for fostering economic stability and boosting international standing. Countries such as Saudi Arabia, Israel, and Japan have crafted strategies that highlight the role of AI in accelerating economic growth and securing a competitive advantage on the global stage.\cite{Galindo2021, JapanAI2022}

In Saudi Arabia, one of the cornerstone initiatives is the "Vision 2030" strategy\cite{SaudiVision2030}, which aims to position the country as a global leader in technology. This strategy includes significant financial investments in the AI sector and envisions the integration of AI into all aspects of life—from healthcare to urban planning.\cite{SDAIA2020_broch} Another example is Israel, which has already established itself as a global leader in AI-related startups. Israel's strategy focuses on supporting this sector through investments in research and education, fostering a business-friendly environment, and encouraging collaboration between the public and private sectors.\cite{IsraelAI2024}

\item \textbf{National Security and Independence.} For some governments, national security and independence are just as important as economic development, particularly when it comes to the implementation of artificial intelligence. 
For instance, countries like China and Brazil place a strong emphasis on strategic autonomy in the development and application of AI, seeking to reduce reliance on foreign technologies and bolster their defense capabilities. They are investing significantly in creating sovereign AI platforms, developing proprietary algorithms, and enhancing cybersecurity to protect critical infrastructure.\cite{ChinaAI2017, BrazilSoverenPlan2024}

\item \textbf{AI Strategies and International Cooperation.}.
The economy of any country is deeply interconnected with its global partners. Therefore, aligning national AI development strategies with international standards goes beyond simply meeting regulatory requirements — it opens doors to economic growth, attracts foreign investments, and builds strategic alliances. For example, the Netherlands and Turkey are shaping their AI policies to align with prominent international recommendations.\cite{Netherlands_GenAI_Vision_2024, TurkeyAI2021}

The Netherlands has aligned its national AI strategy with the EU AI Act, emphasizing transparency, accountability, and the ethical use of AI. This alignment not only ensures compliance with EU regulations but also enhances the country’s attractiveness for international investments and partnerships in the AI sector.\cite{Netherlands_AI_Strategy_2019} Similarly, Turkey leverages its G20 membership and economic ties with the EU to incorporate international AI standards into its national framework.\cite{TurkeyAI2021}

\end{itemize}

Similar to the broader context of AI development strategies, the creation and regulation of large language models (LLMs) are often influenced by diverse strategic and political goals. Several primary motivations can be identified:

\begin{itemize}
    \item \textbf{Data privacy and national sovereignty.} For instance, South Korea and Thailand emphasize the need for sovereign generative AI technologies that align with national interests and legal frameworks.\cite{AIThailand2022} 
    Thailand aims to develop LLMs operating within its borders to safeguard sensitive information from being exported abroad and to maintain adherence to local laws.\cite{Yuenyong2024} In South Korea, a similar strategy reinforces national security while fostering public confidence in AI systems by ensuring data is processed under local legal standards. The citizens can be assured that their data is handled following domestic laws.

    \item \textbf{Geopolitical Concerns.}  
    Geopolitical considerations play a significant role in the development of LLMs, particularly in countries with global or regional strategic ambitions. For instance, nations like China and Brazil view LLMs as powerful tools for managing information flows, maintaining societal stability, and preserving national narratives.\cite{ChinaAIPress2024, brandao_brazil_2024}

    In China, the government exercises strict oversight of language models to ensure they align with socialist ideology. A dedicated state commission evaluates these models, ensuring their content supports socialist values and reinforces official political messaging.\cite{Sheehan2023} These systems are also used to amplify state-approved narratives both domestically and internationally, bolstering geopolitical influence and promoting stability.\cite{Sheehan2023}
    Brazil follows a similar approach, aiming to harness language models for internal security while reducing reliance on foreign-developed systems. By fostering domestic innovation, Brazil seeks to strengthen its technological independence and enhance national security.\cite{brandao_brazil_2024}

    \item \textbf{Safeguarding Cultural Identity Through Language.}

    Israeli tech startup AI21 Labs, founded in 2017, has made a name for itself globally with language models like Jamba and Jurassic.\cite{ai21_about, ai21_jamba} Despite operating in a highly competitive global market, the company has maintained a patriotic mission by incorporating Hebrew into its language models.  This effort highlights a focus on cultural preservation rather than solely financial gains.
    The inclusion of Hebrew was not just a technical achievement but a deliberate effort to strengthen communication in the language and ensure its relevance in modern technology.\cite{orbach2024ai21} This focus on supporting local culture is emblematic of Israeli tech companies' broader efforts to balance cutting-edge innovation with local context.\cite{Newxel2023}
    \item \textbf{Driving Economic Growth for Regional Markets.}
    Latin America and Southeast Asia exemplify how the strategic development of localized LLMs can promote economic progress while addressing market gaps left by global tech corporations.

    The Spanish-speaking population in Latin America presents a vast and largely untapped market for language-focused AI technologies.\cite{grandview2025latin} By producing AI solutions tailored to local needs, regional startups aim to prevent tech giants from monopolizing this market with their Spanish-language models and reaping disproportionate profits.\cite{kim2024latin, sifted2024spain} This approach not only strengthens the region’s tech industry but also ensures that solutions are culturally and contextually relevant.

    In Southeast Asia, the SEA-LION initiative represents a coordinated effort to support the region’s unique linguistic landscape by creating AI models that reflect its cultural diversity.\cite{sea_lion_ai} Locally driven AI innovation stimulates the economy through job creation in tech and enables startups to contribute to the global AI ecosystem. Such initiatives are closely tied to national strategies for long-term economic and technological growth.\cite{imda2023sg}
\end{itemize}

With considerable time since the introduction of national AI strategies and plans, it is now possible to evaluate their progress, impact, and outcomes at multiple levels. In many cases, countries have reexamined and refined their approaches, implemented changes to improve the achievement of original goals, realigned priorities and adjusted their overall vision. The following examples illustrate this process:

\begin{itemize}
    \item \textbf{Drastic Changes in Strategy and Approach.}
Facing unsatisfactory outcomes from its initial AI initiatives, Argentina has made a decisive shift in its strategic direction. Early efforts centered on tight regulations and a closed data ecosystem.\cite{OECD2021} However, as the country’s AI sector began to fall behind other Latin American nations,\cite{latam_2024} the government acknowledged the need for change. Under President Javier Milei, Argentina adopts a more open data policy to attract foreign investment and drive innovation. The goal is to transform Argentina into the "fourth global AI hub" by fostering a business-friendly regulatory framework supporting global tech leaders and domestic enterprises.\cite{hamilton_2024}

    \item \textbf{Refocusing Strategic Goals and Priorities.}
    South Korea’s meticulous evaluation of its AI landscape uncovered critical deficiencies, including a scarcity of quality data and restrictive regulations. These factors contributed to the absence of a high-caliber Korean language model despite the country’s significant technological strengths.\cite{Kang2024}
    To address these issues, the government has revised its approach by boosting financial support for AI-related initiatives, encouraging the development of Korean-language datasets, and adopting a more hands-on approach to AI advancement. In generative AI, the strategy now emphasizes creating efficient small and medium-scale models.\cite{koreapro2024south, eum2024korea}

    \item \textbf{Correction of Timelines and Success Criterias.}
    Turkey’s 2021 National AI Strategy aimed to elevate the country into the global top 20 in AI by 2024, focusing on education, research, and infrastructure. However, the initial strategy lacked precise timelines and measurable indicators to assess development progress.\cite{TurkeyAI2021, Can2022}
    In response to rapid global technological advancements and new regulatory frameworks, Turkey revised its AI strategy. These updates revealed that the initial objectives were misaligned with emerging realities and highlighted gaps in implementation, project delays, and a lack of measurable outcomes. The absence of clear success metrics made it challenging to track progress, emphasizing the importance of transparent performance indicators for accountability.\cite{Can2022, moroglu2024ai_developments}
    The revised 2024–2025 action plan comprises 71 initiatives, including developing Turkish-language AI models, advancements in high-performance computing, and improved data governance. A strong educational focus complements increased funding for innovation and partnerships with local companies to prepare AI experts and expand the country’s workforce in the field.\cite{cbddo2024uyzs, turkey_ai_strategy_2024_2025}
\end{itemize}

These examples clearly illustrate that national AI strategies must be periodically revised to keep pace with rapid technological advancements and to address previous shortcomings. The experiences of Argentina, South Korea, and Turkey, as described above, highlight the importance of setting clear goals, establishing realistic timelines, and demonstrating the flexibility to adapt to new challenges. By examining past missteps, drawing inspiration from successful models, tailoring approaches to the local context, and prioritizing current needs, nations can effectively drive technological innovation, avoid common pitfalls, and create their LLMs.

\subsection {LLM Development.}

\label{subsec:ModelDevelopment}

Developing a Large Language Model (LLM) is a complex and resource-intensive process requiring the integration of multiple components. This process consists of the following essential steps:

\textbf{Data.}
To begin with, the backbone of any LLM is its data. This data, which includes massive collections of text from various domains, must be carefully curated and preprocessed. Handling such volumes — from terabytes to petabytes — requires advanced data pipelines, considerable computational resources, and time to prepare for training.

\textbf{Computational Power.} 
Small models can operate using standard graphics processors, but building large language models necessitates powerful supercomputers or cloud platforms like AWS, Google Cloud, or Azure. These platforms offer GPUs and tensor accelerators capable of processing billions of parameters. However, training such models is a lengthy and costly process, involving significant expenses for hardware and energy resources.

\textbf{Human Resources.}
Building an LLM demands a highly qualified team. Data engineers manage data preparation, machine learning specialists create model architectures, and developers implement the software. The process relies on a synergy of expertise, often drawing on collaboration between academic researchers and experienced professionals in the tech industry to align theoretical innovation with real-world needs.

\textbf{Financial Investments.}
Besides expenses for high-performance computing resources, significant funds are needed for data mining, salaries, and other core expenses. Building a model of this scale often requires investments amounting to tens of millions of dollars, underscoring the capital-heavy nature of these efforts.

\textbf{Additional Tools.} 
 Essential tools such as tokenizers simplify raw text by segmenting it into tokens — units that can be words, parts of words, or characters. This segmentation process ensures that the text becomes accessible to the model for training. A well-designed tokenizer not only ensures precision but also optimizes performance by reducing the number of tokens, which directly lowers computational demands and costs. \cite{shivanandhan2024tokenizers}

In this section, we will provide examples illustrating the process of creating language models in various countries worldwide. First, we will begin with a detailed case study of building a language model in subsection \ref{subsubs:example}. Then, in subsection \ref{subsubs:government}, we will discuss approaches to implementing LLM development projects funded by governments. Finally, in subsection \ref{subsubs:private}, we will examine models initiated by private businesses.

\subsubsection{LLM Development Example}
\label{subsubs:example}
In this subsection, we will explore the creation of language models through the example of Fugaku-LLM (Japan, 2024), illustrating how the development and coordination of language models can unfold.
The Fugaku-LLM project is designed for research and commercial use, focusing on enhancing the processing and generation of Japanese text. It was launched as part of Japan's broader strategy to strengthen AI research capabilities, utilizing the powerful Fugaku supercomputer.
Funding: Although specific funding details are not provided, the project involved substantial contributions from participating organizations, along with government backing through Japan's policy initiatives focused on utilizing Fugaku. \cite{fujitsu2023llm}

With 13 billion parameters, Fugaku-LLM surpasses many other models developed in Japan, which typically have around 7 billion parameters.
The model was trained on the RIKEN Fugaku supercomputer, one of the world's most powerful computing systems. Fugaku’s extensive parallel computing environment enabled efficient training, with 13,824 of its 158,976 nodes (individual components working together for large-scale computations) utilized, highlighting the enormous computational resources dedicated to the project. 
The model’s training data includes around 380 billion tokens, with roughly 60\% in Japanese. This data spans multiple fields to improve the model’s flexibility for tasks like generating natural dialogues and processing text.

Overall, Fugaku-LLM is a collaborative effort between several organizations, each contributing to a specific part of the development process. Here are the main participants and the roles they played: \cite{fujitsu_fugaku_llm_2024}
    
\begin{itemize}
\item  \textbf{Tokyo Institute of Technology}, a leading technical university in Japan, played a pivotal role in overseeing the Fugaku-LLM project. The institute was responsible for optimizing the neural network's architecture and ensuring its smooth integration into Fugaku's distributed computing environment. This involved balancing the trade-off between computational efficiency and model accuracy to ensure the best possible performance. \cite{fujitsu_fugaku_llm_2024, tokyo_tech_overview}

\item  \textbf{Tohoku University,}, recognized for its contributions to science and engineering, took responsibility for data processing. The team concentrated on gathering diverse, high-quality datasets, adapting them to the specific requirements of large-scale language model training, and creating efficient, scalable data processing pipelines to handle vast quantities of data. \cite{fujitsu_fugaku_llm_2024, tohoku_university_overview}

\item \textbf{Fujitsu Limited}, a global leader in technology and co-developer of the Fugaku supercomputer, made a crucial contribution to the computational framework of the project. The company concentrated on optimizing the training workflows, developing efficient parallelization techniques, and fine-tuning system configurations to maximize the throughput of Fugaku’s nodes. Their experience ensured the efficient harnessing of Fugaku’s enormous computational capacity for large-scale, effective training on this infrastructure. \cite{fujitsu_fugaku_llm_2024, fujitsu_about}

\item  \textbf{RIKEN}, as Japan's premier research institution and a key contributor to the Fugaku project, they played a pivotal role in overseeing the distributed training architecture. The team focused on strategies to minimize communication overhead between nodes, improving multi-node operation efficiency and enabling the training of extremely large models without major bottlenecks. \cite{fujitsu_fugaku_llm_2024, riken_about}

\item \textbf{Nagoya University} brought its experience in applied artificial intelligence to the project. The team examined the use of Fugaku-LLM in scientific and technical contexts, concentrating on tasks performed by generative AI. Their contribution was to explore potential use cases of the model beyond traditional language tasks, especially in cases requiring adaptation to specific domains. \cite{fujitsu_fugaku_llm_2024}

\item  \textbf{CyberAgent Inc.}, a top technology company in Japan, offered access to unique Japanese-language datasets. This greatly enhanced the training corpus used for the model, improving its capacity to generate and understand the subtleties of Japanese text. Their contribution ensured that the model would be of high quality for practical deployment in real-world applications for Japanese language. \cite{fujitsu_fugaku_llm_2024, cyberagent_ai}

\item \textbf{Kotoba Technologies Inc.}, a Japanese company and a newcomer in the AI and natural language processing field, was responsible for adapting deep learning frameworks for the Fugaku architecture. The team optimized code performance for the supercomputer's unique hardware, ensuring that model training and data processing were done as efficiently as possible. \cite{fujitsu_fugaku_llm_2024, kotoba_tech_home}
\end{itemize}
    
    \subsubsection{State-Funded Models}
    \label{subsubs:government}

As previously stated, large language model training is a highly complex and resource-demanding process, requiring coordination and collaboration among various organizations. From the examples provided earlier, it is evident that the successful development of an LLM requires the involvement of a diverse set of institutions.

When the government is the driving force behind creating an LLM, it is vital to involve academic institutions, AI-focused private companies, and other technology and data-related organizations. This approach ensures that the necessary resources, expertise, and infrastructure are available for the successful implementation of the project.

In general, different approaches can be distinguished in how institutions or organizations are involved in the creation of sovereign models. One approach is the close collaboration of various organizations and institutions from both the public and private sectors. Another option is to involve only government organizations and some parts of the academic community, with responsibilities divided across different areas of development. The last approach is the creation of a dedicated organization responsible for the overall model development. Below, we will look at a few additional examples from countries that have adopted these approaches.

\textbf{GPT-NL(The Netherlands)}

The Netherlands is one of the countries that has declared plans to create and fund a sovereign language model. In this case, the plan involves three non-profit organizations connected to the government: TNO (Netherlands Organization for Applied Scientific Research), NFI (Netherlands Forensic Institute), and SURF. Each of these organizations will be responsible for different aspects of the model creation. \cite{barbereau2024gptnl}

TNO is a research organization with over 3,000 professionals, specializing in applied research in various sectors, including technology and artificial intelligence. In the GPT-NL project, this organization would playing a key role by offering expertise in AI and data processing for the development of the language model \cite{tno_profile}.

NFI focuses on forensic research, ensuring transparency and the ethical use of language models such as GPT-NL. The institute brings its expertise in forensic analytics to ensure that the model's output is trustworthy \cite{nfi}.

SURF is a cooperative of Dutch universities and research institutions that offers IT infrastructure and technical support. In the GPT-NL project, SURF contributes by providing the technological infrastructure and facilitating collaboration among partners to integrate the model into R\&D \cite{surf}.

\textbf{GPT-SW3 (2023, Sweden)
}
    Another example is the development of GPT-SW3, a large-scale generative language model for Scandinavian languages, where several key organizations and corporations were involved. 
    The training of the earlier 3.5B version of GPT-SW3 used 16 nodes over 2.5 days. For the larger 175B version, it is estimated that would require 30 nodes and approximately 90 days \cite{sahlgren2022gpt}.
    
    Key participants and sponsors:
\begin{itemize}
    \item \textbf{AI Sweden.} The development of GPT-SW3 was led by the NLU research group at AI Sweden. They played a key role in coordinating the project and ensuring its successful development \cite{a2025_swedish}.

    \item \textbf{RISE.}  A Swedish research institute that worked alongside AI Sweden on the GPT-SW3 project. Their participation was vital in offering additional expertise and resources \cite{rise_about, wasp2023gptsw3}.
    
    \item \textbf{WARA.} A part of the Wallenberg AI, Autonomous Systems, and Software Program (WASP), WARA contributed to the project with a focus on media and language aspects. Their involvement assisted in validating and improving the model’s capabilities for both general tasks and more specific ones, such as those in the media field \cite{wasp2023gptsw3}.
    
    \item \textbf{Linköping University.} The training of GPT-SW3 was carried out on Linköping University's Berzelius supercomputer.
    
    \item \textbf{Nvidia}. While Nvidia isn’t explicitly listed as a co-author in some sources, and its involvement in the project isn’t fully outlined, the developers mention that they used Nvidia’s NeMo Megatron framework for training GPT-SW3 \cite{wasp2023gptsw3}.
\end{itemize}

\textbf{Arabian LLM(2024, Saudi Arabia)}

ALLaM is a set of large language models created to advance Arabic Language Technologies (ALT). These models are designed to achieve a high level of proficiency in both Arabic and English \cite{bari2024allam}. The primary force behind ALLaM’s development is the Saudi Data and Artificial Intelligence Authority (SDAIA), a government entity established and funded as part of Saudi Arabia’s sovereign AI development strategy \cite{sdaia_2023_about}. SDAIA has also partnered with IBM, facilitating the integration of ALLaM into the IBM WatsonX platform. This partnership ensures that both public and commercial users can access ALLaM, enabling widespread applications across various domains \cite{sdaias_allam_2024}.

The ALLaM models were trained on a massive dataset consisting of over 3 trillion tokens in both Arabic and English. This dataset was compiled with the help of more than 400 experts and 160 government organizations, making it the largest Arabic-language dataset in the world \cite{sdaia2024global}. ALLaM’s architecture is based on the Llama-2 model (an open-source model developed by Meta), with some models initialized from Llama-2 and others trained from scratch. The training process utilized the NVIDIA/MegatronLM infrastructure, allowing for efficient scaling of large language model training \cite{bari2024allam}.

ALLaM has achieved exceptional results in both Arabic and English language tasks, making it a versatile tool for various applications. The model has topped several key Arabic benchmarks, including MMLU Arabic, ACVA, and Arabic Exams. Another important feature is its open licensing under the SDAIA License, which allows both commercial and governmental entities to use and adapt the model as needed \cite{bari2024allam}.

In the future, SDAIA intends to further expand the ALLaM model family by continuing to train models based on current open models and creating entirely new architectures. Additionally, ALLaM’s availability opens up significant research potential in the field of Arabic NLP, especially in developing models that better capture the intricacies of Arabic dialects \cite{crestana2024sdaia, sdaia2024global}.

\subsubsection{Models Developed by Private Initiatives}
\label{subsubs:private}

When the creation of local language models is initiated by a private organization, it is usually led by large technology companies. However, ambitious startups may occasionally take part, but those cases are observed only under specific circumstances. These organizations generally focus on widely spoken languages with substantial commercial potential to make profits and gain market share.
Such models can vary significantly in both form and resources, as well as in their approaches to distribution. Some companies take an open-source approach, releasing their models publicly so the community can use and refine them. Others offer access to their models through paid APIs (in this context, sets of guidelines and rules that allow interaction with the model), monetizing the technology. Additionally, some companies build entire platforms around their language models, providing integrated solutions tailored to the needs of businesses and individual users.

Here are some more examples of LLM development that are relevant in this context.
\begin{itemize}

\item \textbf{QWEN (China, 2023-2024)}

The \textbf{Qwen} model family consists of large language models (LLMs) developed by Alibaba Cloud, a division of the Chinese technology giant Alibaba Group \cite{qwen_github}. Positioned as a leading player in the competitive language model market, Qwen reflects Alibaba’s ambition to lead AI innovation both in China and globally. The model's name, meaning "a thousand questions," reflects its purpose of addressing a broad spectrum of user queries \cite{mah2024alibaba_ai_models}.

Qwen was developed as part of Alibaba's broader AI strategy, leveraging the company’s extensive expertise in cloud infrastructure, data management, and AI research. The training process took place on Alibaba Cloud’s cutting-edge computing platforms, utilizing their hardware and scalable cloud resources. Although the technical specifics vary across different iterations of Qwen, the model's multilingual capabilities, particularly its focus on Chinese and English, are a central feature \cite{shao2024qwen}.

Alibaba presents Qwen as a flexible and scalable AI solution tailored to diverse industries, including e-commerce, customer support, finance, education, and healthcare. Seamlessly integrated into Alibaba’s broader ecosystem, Qwen empowers businesses with tools for enhanced customer engagement, personalized recommendations, and data-driven decision-making. Beyond internal applications, Qwen is also offered as an AI-as-a-Service (AIaaS) product, extending its reach to a broader audience \cite{alibaba2023qwen}.

Recognized as a leader in China’s AI sector, Qwen is gradually expanding its footprint in global markets. Its adoption spans startups to large enterprises, all leveraging its capabilities to boost efficiency and improve client interaction. To encourage further development, Alibaba has open-sourced select versions of Qwen, empowering the global AI community to customize and innovate with its technology \cite{alibaba2024qwen}.
Globally, Qwen competes with AI heavyweights like OpenAI’s GPT, Anthropic’s Claude, and Google’s Gemini. While it is still catching up in terms of global visibility, Qwen leverages Alibaba’s comprehensive cloud infrastructure, massive datasets, and strong market influence across Asia \cite{definition2023most, techinasia_chinese_ai_models}. This combination makes Qwen more than just an AI tool — it’s a cornerstone of Alibaba’s strategy to lead in the realms of cloud computing and AI innovation.

    \item \textbf{Sabia models (Brazil, 2023-2024)}

    Maritaca AI, a Brazilian startup, specializes in developing artificial intelligence for natural language processing with a strong focus on the Portuguese-speaking market. Combining scientific research with commercial development, they create high-quality products that meet local demands. Their Sabiá series of language models provides an API for integrating powerful AI tools into business applications, charging only for the resources used\cite{unicamp2024chatgpt_dalle}.
    The latest model, Sabiá-3, has shown excellent performance in national exams and is significantly more affordable than its competitors. The previous version, Sabiá-2, outperformed GPT-3.5 in most tests and demonstrated similar performance to GPT-4 in other tasks.\cite{maritaca_ai, tsar2024proceedings} They also developed the Sabia-7B model, optimized for efficient use on low-power platforms.\cite{maritaca2023sabia}
    Maritaca AI differentiates itself from global leaders with its focus on the Portuguese market, competitive pricing and flexible terms, making it a desirable option for various sectors, including education, business, and government initiatives \cite{maritaca2023sabia}.

    \item  \textbf{HyperCLOVA (South Korea, 2024)} is a language model developed by Naver, a major corporation that focuses on AI technology for the global and Korean market and operates the country’s leading search engine \cite{naver_corporation, statista2024search}. Their model is designed to excel at understanding Korean, enabling it to work effectively with local text and business processes \cite{naver2023hyperclova}. 
    The development of HyperCLOVA involved over 300 experts from various Nexus divisions, illustrating the company’s vast efforts \cite{yoo2024hyperclova}. Naver is one of South Korea’s leading corporations, heavily investing in AI, with over\$700 million dedicated to AI development, showcasing its substantial presence in the market \cite{naver2024ai_investment}. 
    The HyperCLOVA model is designed to process large datasets and can be applied in various fields, including finance, education, and order management. It is integrated into businesses through strategic partnerships \cite{naver2023hyperclova_korea}.
    HyperCLOVA’s standout feature is its specific optimization for the Korean language, offering an enhanced user experience. Additionally, the model adheres to stringent security and ethical standards, ensuring safe usage \cite{yoo2024hyperclova}. Naver has partnered with innovative startups to create new AI solutions on the HyperCLOVA platform, further contributing to the region's AI landscape \cite{clova_studio}.

\end{itemize}

\subsection{Investment}
\label{subsec:ModelInvestments}

Over the past few years, governments in many countries have stepped up their investment in artificial intelligence, with a particular emphasis on the creation of sovereign large language models. These efforts, aimed at advancing scientific discovery, economic development and securing technological independence, encompass funding research projects, building state-of-the-art infrastructure, and implementing programs that encourage innovation in language technology. In that way, nations are solidifying their standing in the global AI ecosystem. In this section, we will examine examples of these investments in subsection \ref{subsubs:general_investments}, their focus areas in subsection \ref{subsubs:trends}, and discuss the sources of funding in subsection \ref{subsubs:sources}.

\subsubsection{AI Funding Examples} \label{subsubs:general_investments}

The following examples illustrate how the European Union and four different nations have been investing in artificial intelligence development in recent years. We will also highlight the scale of these investments to showcase the level of engagement in AI advancement outside the leading players such as the United States, China, and France.

\begin{itemize}

    \item \textbf{European Union.}
    The European Commission is investing €1B annually into AI projects between 2024 and 2027, with funding directed toward Horizon Europe and Digital Europe programs \cite{eu_ai_strategy}. Horizon Europe has allocated \$50M of this funding specifically for initiatives focused on large language models \cite{ec2024new}. Furthermore, through the Horizon Europe framework, EuroHPC is managing the financing and establishment of seven "AI Factories," state-of-the-art supercomputers backed by a €1.5B investment \cite{ec_ai_factories}. (See subsection \ref{subs:intr_programs_eu} for further information.)

    \item \textbf{Sweden.}
    Vinnova, the Swedish Innovation Agency, has invested €135 million in AI-related projects from 2020 to 2024, including \$585,838 allocated to training the GPT-SW3 model \cite{vinnova2024gpt}. This initiative has established a collaborative ecosystem for AI and language model research within the country. By 2025, Vinnova intends to increase its investment by an additional \$0.9M to bolster infrastructure and advance AI capabilities \cite{vinnova2023nextgen}.
    Furthermore, Sweden’s government has committed to raising its funding for research and innovation to 6.5B SEK (equivalent to €350M) by 2028. This expanded financial support aims to accelerate the development of emerging technologies like AI, enabling Sweden to remain competitive on the global stage \cite{govse2024research}.

    \item \textbf{The Netherlands.}
     Annual funding for AI research and innovation in the Netherlands stands at roughly €45 M \cite{ec2022netherlands}. The Ministry of Education, Culture, and Science has contributed €18 million to build a new supercomputer for SURF, while the Ministry of Economic Affairs is supporting the GPT-NL project with €13.5M. GPT-NL is designed as an open large language model aligned with Dutch cultural values and standards \cite{Netherlands_AI_Strategy_2019}. Additionally, the Dutch National Growth Fund has allocated €204.5 million to strengthen the nation's AI sector through the AINEd initiative \cite{nlaic_ained}.

    \item \textbf{Brazil.}
    Brazil is investing 23 billion Brazilian reals (approximately \$4.2 B) over four years as part of its national AI development plan \cite{forbes2024brazil}. This plan includes a \$40 M allocation to establish legal frameworks for AI and \$200 M specifically dedicated to developing a sovereign language model and the necessary infrastructure \cite{mcti2021ebia}.

    \item \textbf{Israel.}
    The Israel government invested \$140M in its National AI Program (2021–2024), focusing on developing infrastructure and founding a National AI Research Institute to enhance academia-industry partnerships \cite{israel_national_ai_program_2024}.
    Within that planning, the government allocated \$1.8M for an association focused on advancing Hebrew and Arabic processing technologies to boost NLP development \cite{aiisrael2024nlp}. An additional \$7M (30 million shekels) has been invested through the Israel Innovation Authority in various projects to improve access to data and conversational AI models for these languages \cite{wrobel2023israel}.
    Furthermore, Israel partnered with NVIDIA in 2023 to develop the Israel-1 supercomputer for large AI models, a project costing several hundred million dollars \cite{wrobel2023nvidia}. Additionally, the Israel Innovation Authority and Ministry of Finance launched a \$160 M fund in 2024 to further promote innovation and to stimulate tech investments, particularly in AI and other emerging technologies \cite{xinhua2024israel_fund}.
\end{itemize}

\subsubsection{Trends in AI Investments} \label{subsubs:trends}

This subsection examines prominent examples of artificial intelligence investments, reflecting broader trends in the field. The analysis is divided into three main categories: investments in talent to bolster the nation’s skilled workforce, infrastructure investments critical to long-term development, and overarching trends in investment growth, demonstrating increased funding well beyond the allocations in original AI plans.

\begin{itemize}
    \item \textbf{Human Capital Investments::}
    The development of AI is driven by the expertise of those designing it. These individuals not only spark innovation but also enable the integration of AI into diverse aspects of life and elevate global competitiveness of their country. Skilled experts serve as the cornerstone of technological sovereignty and economic progress, actively participating in the creation of AI technologies. That is the reason why investing in talent is crucial for ensuring the sustainable growth of artificial intelligence.
    Many national strategies set specific targets, such as training N AI engineers, hiring M AI developers, or establishing K R\&D centers focused on AI. Implementing these goals requires a multifaceted approach, often making it difficult to determine precise funding levels for these initiatives. Nevertheless, here are some illustrative cases:
    \begin{itemize}
        \item \textbf{Thailand} plans to invest 1 billion baht (\$28M) to train 30,000 AI professionals under its national strategy, representing a substantial portion of the overall \$42M budget \cite{aithailand2023annual, vietnamvietnamplus_2024_thailand}.
        \item In the second phase of \textbf{Israel’s} national strategy, which is funded with approximately \$133M, a strong focus is placed on advancing research and innovation centers as well as the development of human capital \cite{innovation_authority2024second}.
        \item Recently, ten universities in \textbf{Japan} received over 100 million yen (\$0.7M) in annual grants aimed at growing AI talent and promoting international collaboration. This initiative highlights Japan’s strategy to strengthen its educational system \cite{nikkei2024japan}.
        \item \textbf{The Saudi Arabian government} has established a variety of programs to stimulate AI startup growth, including the "Garage" incubator at the King Abdulaziz City for Science and Technology (KACST), the "Taqadam" accelerator at King Abdullah University of Science and Technology (KAUST), GAIA by the Saudi Data and AI Authority (SDAIA), and the National Technology Development Program (NTDP). GAIA specifically targets early-stage generative AI initiatives. Collectively, over \$200M has been dedicated to these efforts \cite{global_ai_summit_state_of_ai_saudi_arabia}.
        \item As part of its national strategy, the \textbf{Brazilian government} has earmarked funds for AI-related education, including 183.24 million Brazilian reals (\$36M) to create 5,000 training opportunities over four years, 152 million reals (\$30M) for doctoral scholarships, and 194.2 million reals (~\$38M) for scholarships for undergraduate and graduate students in Brazil. These initiatives are designed to enhance the AI skills of 100,000 Brazilians \cite{mcti2024ia}.
        \end{itemize}

    \item \textbf{Infrastructure Investments:}
    Training large AI systems that require massive computational power often leads companies and research centers to rent resources from cloud providers or rely on international data centers. However, this approach has its drawbacks: it limits control over data, increases long-term expenses, and creates dependence on external providers. Therefore, funding for proprietary computing resources becomes a strategic investment, unlocking numerous possibilities, such as the training of sovereign language models.
    
    As mentioned previously, the Netherlands and Israel have allocated tens of millions of dollars to develop supercomputers for AI, particularly for the purpose of training their own large language models. Other nations are also making substantial investments into computational infrastructure:

    The \textbf{Brazilian government} is set to invest more than \$18 million in the Santos Dumont supercomputer, which is expected to rank among the top five supercomputers worldwide. This investment will expand AI research capabilities, particularly for the development of a sovereign language model \cite{mcti2024santos_dumont}. In addition, in September 2024, Petrobras (the state-owned oil company) announced plans to invest about \$90 million in acquiring five new Lenovo supercomputers for its research and innovation center in Rio de Janeiro \cite{brazilenergy2024petrobras}.

    \textbf{In Israel}, another supercomputer is planned to be built for training large AI models, with a tender already announced for \$64 million \cite{wrobel2024israel_ai_supercomputing}.

    The \textbf{Indian government} has approved an investment of 103 billion rupees (1.24 billion USD) to enhance the national artificial intelligence infrastructure. A major part of this initiative is the plan to build a supercomputer featuring at least 10,000 GPUs, which will be integrated into the "IndiaAI Compute Capacity" project \cite{dunn2024india}.

    \textbf{South Korea} is set to create a dedicated fund of 1.4 trillion won (~\$1B) to support innovative semiconductor firms operating in the AI sector. The President noted that the success of the semiconductor industry is deeply connected to advancements in AI, with the aim of making South Korea one of the top three global AI leaders by 2027, alongside the US and China. The government also plans to establish national AI computing centers to enhance supercomputing infrastructure and hopes to achieve major breakthroughs in AI technology \cite{kang2024south}.

    \textbf{Foreign Funding}

    Tech giants like NVIDIA, IBM, and others are also contributing to infrastructure investment, and not just in leading AI countries. For example:

    In June 2024, \textbf{Microsoft} committed \$3.2 billion to the development of AI and cloud technologies in Sweden. This investment will add 20,000 GPUs to data centers and train 250,000 individuals in basic AI skills. The choice of Sweden is partially due to its abundance of green energy resources \cite{fortune2024microsoft_swedish_investment}.
        
    As another example, \textbf{Google} allocated 36 billion Thai Baht (approximately \$1B) for the construction of a new data center and the expansion of cloud infrastructure in Thailand, specifically in Chonburi and Bangkok. This is part of a broader Google strategy to enhance its AI capabilities in the face of rising competition from other tech giants such as Microsoft and OpenAI \cite{reuters2024google_thailand}.
    
    More details about such collaborations can be found in section \ref{sec:BigCompanies}.

    \item \textbf{Investments Growth.} 
    Another trend that can be observed is the increase in government funding for AI, beyond what was initially planned within national strategies. Many countries now recognize that their early plans often failed to account for the rapid pace of AI development, and they see additional investments as crucial to keeping up in this area. Some examples demonstrating this trend include:

    \textbf{Singapore}
    is increasing its funding for AI applications in science. Through the National Research Foundation (NRF), the country has allocated S$120 million ($88M) specifically for the "AI for Science" initiative. This grant supports projects that apply AI in scientific research, in addition to the more than S$500 million ($370M) already invested since 2019 under the national AI Singapore program \cite{smartnation2024}.

    \textbf{In Sweden}, by the end of November 2024, the government established Artificial Intelligence Commission announced that the government should increase AI investments by 12.5 billion SEK (~\$1.14B) over the next five years to ensure the country becomes a major player in the AI field and stays on par with other nations \cite{ericsson2024ai_commission}.
    
\end{itemize}

\subsubsection{Funding sources for LLM development} \label{subsubs:sources}

In this section, we will examine how investments in artificial intelligence are formed, drawing on specific examples. The section will be divided into four parts, addressing three main approaches: investments from the state budget, contributions from the private sector and international grant funding.

\begin{itemize}
    \item \textbf{State-backed}

    The government plays a key role in initiating and sponsoring the development of sovereign LLMs. In most cases, governments publicly commit to the creation and funding of these models as part of their national strategies. For example, Brazil has allocated \$200M for its language model development and the associated infrastructure, while the Netherlands has invested €13.5M in GPT-NL. \cite{mcti2021ebia, Netherlands_AI_Strategy_2019}

    In Singapore, state-funded institutions are developing large language models as part of the country’s AI strategy. \$52M has been set aside for the development of SEA-LION (Southeast Asian Languages In One Network), an open-source family of models supporting 11 regional languages developed by AI Singapore. \cite{govinsider2023singapore, aisingapore_sealion}

    In Thailand, the second phase of the national AI strategy prioritizes the development of the Thai large language model (ThaiLLM). Approximately 120 million baht (\$2.5M) were designated for its core development and the creation of chatbots for sectors like healthcare and tourism. \cite{vietnamvietnamplus_2024_thailand}
 \end{itemize}

  \begin{itemize}  
    \item \textbf{Business}

    There are several ways in which businesses can finance the creation of language models. One method is through investment in the company’s own R\&D center dedicated to creating models that support the national language. While the development budget of such models is usually kept confidential, these models are often released publicly, making them available for use by others.

    Samsung’s Research and Development Center is dedicated to creating technologies of the future that benefit society through continuous innovation and intelligent solutions. In 2021, the company released the KoreAlBERT language model, specifically tailored to improve Korean language processing. \cite{lee2021korealbert} In addition, Samsung developed the generative language models Samsung Gauss and Gauss2, which aim to optimize the size and performance of models for specific Samsung products and services, supporting multiple languages, including Korean. \cite{samsung_research_ai}

    Baidu, the leading Chinese technology giant, has developed a series of ERNIE (Enhanced Representation through Knowledge Integration) models as part of its large-scale research and development initiatives, aiming to enhance language understanding and generation for Chinese and other languages. \cite{baidu2023ernie} This open-source model supports a broad range of natural language processing applications, including question-answering and sentiment analysis. \cite{huggingface2023ernie} ERNIE was designed to advance AI capabilities and compete with global leaders like OpenAI, with a particular focus on optimizing Chinese language processing. The company is set to release an improved fifth version of its model in 2025. \cite{baidu_wenxin_llm_2024}
    
     The second strategy is investing in a company that develops language models. This usually applies to large or foundational models with widespread potential applications. For instance, Mistral models – a family of large language models developed by researchers in France during 2023-2024, most of which support multiple languages (including English, French, Spanish, German, and Italian). \cite{mistral_ai_technology} The company secured over \$1.1B through two investment rounds from a variety of companies and funds (including Lightspeed Venture Partners, Andreessen Horowitz, Nvidia, Samsung Venture Investment Corporation, Salesforce Ventures, BNP Paribas, Cisco, IBM, Sanabil Investments, and many others). \cite{ain_mistral_ai_funding}

     The third type of funding focuses on language models designed for specific needs.

     Saudi Aramco, the world’s largest oil company based in Saudi Arabia, needed a large language model (LLM) to support its business functions. To meet this need, they collaborated with the Saudi Accelerated Innovation Laboratory (SAIL) – a digital innovation hub that works with a broad network of companies and startups to create advanced digital solutions. Aramco financed the development of its proprietary language model. \cite{lucidity2024aramco} In March 2024, Aramco and SAIL introduced METABRAIN – a generative AI model designed for industrial-scale applications aimed at optimizing business processes. With 250 billion parameters, METABRAIN is the largest industrial AI model to date. It was trained on trillions of words sourced from both public data and Aramco’s 90-year-old internal archives. The model improves extraction operations, forecasts oil product prices, market behavior, and geopolitical trends. \cite{malin2024aramco} In general, METABRAIN is a crucial tool in Aramco’s digital transformation, boosting productivity and driving innovation across its operations.

    \item \textbf{Grants}
        
    EU grant programs such as CLARIN and Horizon Europe (outlined in more detail in section \ref{sec:eu_programs}) provide funding for projects aimed at developing language processing tools. 
    For example, Turkey received approximately \$14M between 2014 and 2020 for 27 language processing projects. \cite{TurkeyAI2021} Poland has been involved in the CLARIN program since 2014. \cite{clarin_pl_about} As part of this program and its investments, several Polish language text and audio corpora were created, \cite{clarin_pl_resources} along with various tools for natural language processing, including CLARIN-PL, a chatbot designed to improve Polish language response generation as an enhancement to ChatGPT. \cite{clarin_pl_chat}

      EuroLLM was awarded a grant for the use of supercomputing resources under the EU’s "Euro HPC for Infrastructure" program and received about €4 million in grant support from the EU’s Horizon Europe Research and Innovation Actions. The aim of the project is to develop a multilingual and multimodal language model that can support all 24 official EU languages. \cite{unifiedtranscriptionandtranslationforextendedreality_2022_transforming, eurollmteam_2024_eurollm9b}
      The development process involved multiple universities and organizations across various countries, including Unbabel (an AI-based translation platform) \cite{unbabel_about}, Instituto Superior Técnico, Instituto de Telecomunicações, University of Edinburgh, Aveni, University of Paris-Saclay, University of Amsterdam, Naver Labs (an R\&D division of the prominent South Korean internet company NAVER) \cite{naver_labs_europe}, and Sorbonne Université. \cite{utter2023eurollm}

    Another example is the Brazilian startup Maritaca AI, which received a \$1 million grant from Google for the use of Tensor Processing Units (TPU) on Google Cloud. The project focuses on developing a large language model (Sabria2/3) specifically designed for the Portuguese language. \cite{maritaca_ai} Despite being much cheaper than OpenAI models, they have managed to achieve results that are on par with, or even exceed, those of OpenAI models. \cite{peq42_maritalk}

    \textbf{Government Academic Grants}. PhD students frequently focus on improving, training, or even creating language models in their research, with such projects often funded by universities through their operational budgets. While it is hard to specify exactly how much grant funding is allocated to academic projects involving language models, it is possible to understand the general funding scale. For example, Poland's National AI Strategy outlines around 20 grant programs for AI innovation projects, totaling \$1 billion in funding. Most of these programs are aimed at supporting businesses, governmental entities, and technological products. A few specifically fund academic and industrial research in AI, with several tens of millions of dollars allocated for these projects. \cite{poland_ai_policy_2020}
        
    \item \textbf{Multiple Funding Channels}

    Often, the development of a language model is financed by a combination of organizations. 

    For example, in the creation of OpenThaiGPT, both public and private entities participated, with funding coming from government agencies, private sponsors, and community-driven initiatives. \cite{openthaigpt, openthaigpt_sponsors} The main organizations involved in the funding included the National Science and Technology Development Agency (NSTDA), more specifically its computational department, ThaiSC, the Thai Artificial Intelligence Entrepreneurs Association (AIEAT), and the Artificial Intelligence Association of Thailand (AIAT), which managed the government involvement, including the use of the LanTa supercomputer for model training. \cite{nstda2022supercomputer}

    Another organization involved was CmKL University, a collaboration between Carnegie Mellon University and King Mongkut’s Institute of Technology Ladkrabang (KMITL). \cite{cmkl_university} CmKL contributed by investing and providing intellectual resources.
     From the business side, Pantip, one of Thailand's leading online discussion platforms, was involved. \cite{semrush2024top} Besides financial sponsorship, Pantip also supplied data for training OpenThaiLLM. \cite{leesa_nguansuk2023open_thaigpt}
    
    A similar example is the Faros project (Greek AI Factory), which received investments from the Greek government and EU grants. With total investments of €30 million, it is one of seven approved AI factories submitted by EU member states to the European Commission. \cite{greeknewsagenda_2024_greeces} Half of the budget, €15 million, is provided by the government, while the remaining funds are supplied through the joint EuroHPC fund. \cite{greeknewsagenda_2024_greeces}

\end{itemize}

\section{How to regulate LLMs?}
AI models can significantly affect society, especially in areas with processing confidential information, decision-making, and national security (for example, they may aid in the spread of misinformation). Equally crucial are the ethical issues stemming from biases inherent in training data. This makes it essential for AI models to be high-quality, fair, and transparent in their operation. Effective regulation of large language models (LLMs) focuses on two main aspects: overseeing their application and regulating their training, with particular emphasis on the selection of training data.

\subsection{LLM Usage Regulations}
Countries adopt different strategies to restrict and regulate the use of language models. Nevertheless, the primary approaches to their regulation can be categorized as follows:

\begin{itemize}
    \item \textbf{Access Restrictions.}
    Countries may impose bans on models that haven't received certification prior to their market launch. In China, for example, large language models are mandated to undergo a state-level review to ensure they align with national policies before being released to the public. This regulatory approach forms part of a broader strategy to guide the development and implementation of artificial intelligence (AI) in ways that reflect national priorities and the socialist ideology.\cite{LuoDan2023} Additionally, China has set strict guidelines for moderating AI-generated content, requiring that such content be clearly marked, while providers are obligated to oversee and filter outputs to prevent the spread of harmful or unlawful materials.\cite{xuan_2024}
    
    \item \textbf{Licensing for Government and "Sensitive" Sectors.}
    For example, in South Korea, a company seeking approval to implement AI solutions, including language models, in critical sectors like government, defense, or other sensitive fields (such as healthcare, justice, etc.) must first secure a government license. This requirement is outlined in the AI Basic Act \cite{AI_Basic_Act_2024}. The law is set to come into effect in January 2026. It will establish a legal framework to ensure the safe development and implementation of AI technologies, with a strong emphasis on risk reduction.\cite{yoon_2024}

    \item \textbf{AI Application Monitoring.}
    In Brazil, Bill No. 2,338/2023\cite{PL_2338_2023} proposes the establishment of a legal framework for AI usage based on a risk-driven regulatory approach. A central element of this system is the introduction of regular audits for AI applications used in government administration, especially those deemed high-risk (e.g., involving access to private personal data used for decision-making, etc.)\cite{White_and_Case_LLP_2024}. The aim is to promote transparency and accountability in governance by making developers and users accountable for AI tools, ultimately reducing risks associated with automated processes in public administration.
        
    This approach is in line with global trends focused on enhancing oversight of artificial intelligence technologies, particularly in light of the need for responsible governance amid rapid technological progress.
    
    In the \textbf{Netherlands}, the regulation of LLMs is evolving as part of the national artificial intelligence strategy \cite{Netherlands_AI_Strategy_2019} and the generative AI plan \cite{Netherlands_GenAI_Vision_2024}, alongside the EU AI Act (European Union Artificial Intelligence Regulation) \cite{EUAIAct2024} (which is discussed further in subsection \ref{sec:AI_Act}). AI regulation in the country is being implemented in stages. The advisory and mandatory sections of the AI Act, which was passed in early April 2024, will become effective on February 2, 2025 \cite{Netherlands_AP_Guidance_2024}.
    These regulations provide the foundation for the responsible development and adoption of AI technologies, balancing safety, ethical concerns, and innovation. For example, starting in August 2027, all AI systems classified as high-risk will have to meet new standards, including requirements for risk management and transparency, and will be required to obtain a license \cite{Netherlands_AI_Document_2023}.
 
    \textbf{Sweden} applies a similar approach in general. The Swedish Commission on Artificial Intelligence, which was tasked by the government to examine the potential applications of AI in different sectors, has recently suggested broadening AI usage in public services, coupled with the introduction of relevant regulations. For example, eSam, a Swedish initiative that includes 34 governmental bodies, is focused on advancing digitalization in public administration while also supervising its progress \cite{Sweden_AI_Commission_2024}.
    The most relevant framework is the AI Use Checklist, which provides guidelines for personal data processing, public service delivery, security issues, government tenders, as well as transparency and explaining decision-making by AI models, and addressing model biases. It is important to note that, like many other regulatory documents, the Checklist often references EU regulations \cite{reichel_2023}.

    \item \textbf{Responsibility Allocation.}
    This principle separates the responsibility between the developers of AI tools and the individuals who use them. For example, in Israel, developers of military AI models are accountable for the security and effectiveness of their systems. However, the Israeli Defense Forces assert that commanders must approve decisions regarding target acquisition, thereby preserving human oversight and accountability in AI-led operations. Thus, AI technologies are viewed as tools under human control \cite{serhan_2024, mimran_dahan_2024}.

\item \textbf{Integrated Risk-Based Approach.}
    A prime example of AI regulation is seen in Thailand, governed by The Royal Decree on Artificial Intelligence System Service Business \cite{draftRoyalDecree2022}. The approach revolves around evaluating the risks that AI technologies may introduce. Primarily, the focus is on potentially harmful or high-risk AI services that could endanger public health, safety, or personal freedoms (for instance, AI used in judicial administration, visa processing, surveillance, or recruitment). The regulatory framework is tied to the level of risk associated with an AI system: unacceptable risks lead to prohibition, high risks require compliance assessment and licensing, while lower risks call for transparency \cite{thai_regularization_state_2024}. 
    Additional regulations elaborate on the criteria and procedures for minimizing potential risks. Furthermore, this regulation bans the development of AI tools that could influence human consciousness, such as AI systems used for automatic ranking or identifying people in public areas and technologies that have access to sensitive data \cite{draftRoyalDecree2022, baker_mckenzie_2023}.
    
    The Office of the National Digital Economy and Society Commission, part of the Ministry of Digital Economy and Society, is responsible for overseeing these regulations. However, there are various nuances to consider, such as the fact that they do not apply to certain institutions or systems that are supervised by specific regulators, like the Bank of Thailand or the Office of the Securities and Exchange Commission \cite{draftRoyalDecree2022}
\end{itemize}
    
\subsection{Who Is Responsible for Regulation?}

In many countries, specialized agencies oversee data protection, ensure privacy, and regulate various aspects of technological activity. These organizations are upholding the rights and safety of citizens while driving the growth of national technological and innovation sectors. They monitor compliance with legislation, regulate the operations of financial institutions, develop policies for artificial intelligence, and establish technical standards. This section dives into the structure and functions of such bodies in different nations.

\subsubsection{Data Protection and Privacy Institutions}
\begin{itemize}
    \item The Netherlands: The Dutch Data Protection Authority (AP) \\
    The Dutch Data Protection Authority (AP) serves as an autonomous oversight body in the Netherlands, tasked with upholding fundamental privacy rights. It ensures that companies and institutions comply with applicable privacy regulations.
    \cite{autoriteitpersoonsgegevens_2023_current}
    
    \item Brazil: National Data Protection Authority (ANPD)\\
    The National Data Protection Authority (ANPD), founded by the General Data Protection Law (LGPD), operates as an independent body with full administrative and budgetary autonomy, enabling it to enforce decisions throughout Brazil. It has the authority to apply penalties for LGPD infractions. The ANPD’s framework includes entities such as the Board of Directors and the National Council on Personal Data and Privacy Protection, responsible for overseeing compliance and issuing regulations.
    \cite{dlapiper_2023_brazil}
    
    \item Colombia: Superintendence of Industry and Commerce (SIC)\\
    The Colombian Data Protection Authority, formally known as the Superintendence of Industry and Commerce (SIC), is empowered to investigate potential violations of data privacy and protection rights for Colombian data subjects based on complaints. SIC can issue binding orders to database controllers, enabling data subjects to access, correct, or delete data collected in violation of their rights. It also promotes educational campaigns on data protection rights. Furthermore, SIC may impose fines following an administrative investigation if a violation is proven.
    \cite{globalcompliancenews_2023_colombia}

    \item Argentina: National Directorate of Personal Data Protection
    The National Directorate of Personal Data Protection in Argentina, which is part of the Argentine Agency for Access to Public Information, serves as the regulatory body overseeing compliance with and enforcement of the Personal Data Protection Law at the national level. Personal data must be processed legally, fairly, and transparently, with clear restrictions on purpose, data minimization, accuracy, retention, and security to safeguard the rights of data subjects. \cite{cookieyes_2023_argentina}
    \cite{cookieyes_2023_argentina}
    
    \item Poland: Personal Data Protection Authority (UODO)\\
    The Personal Data Protection Authority is the national body responsible for data protection in Poland. Located in Warsaw, it is tasked with ensuring compliance with the GDPR in Poland.
    \cite{uodo_2023_edpb}

\end{itemize}

\subsubsection{Authorities Allocating Funding for Institutions and Research}
\begin{itemize}
    \item Netherlands: De Nederlandsche Bank (DNB)\\
    De Nederlandsche Bank is a public limited company fully owned by the state of the Netherlands, acting as the central bank within the European System of Central Banks. DNB collaborates with European colleagues on monetary policy, payment systems, and banking supervision within the Single Supervisory Mechanism. The bank’s management structure consists of the Executive Board, the Supervisory Board, and the National Council, ensuring accountability and oversight of its operations.
    \cite{dnb_2023_organisation}
    
    \item Brazil: Ministry of Science, Technology, Innovation, and Communications\\
    The Ministry of Science, Technology, Innovation, and Communications oversees federal policies in science and technology in Brazil. Along with its research units and affiliated institutions, it operates across the country, funding scientific research, advancing innovation, and driving initiatives aimed at solving national issues through science, technology, and innovation. \cite{govbr_2023_science_technology_innovation}

    \cite{govbr_2023_science_technology_innovation}
    
\end{itemize}

\subsubsection{Authorities,Responsible For Developing AI Policies and Strategies}
\begin{itemize}
    \item Turkey: Digital Transformation Office\\
    The Digital Transformation Office was established in 2018 to ensure a unified strategic approach to the country’s digital evolution. Since its inception, the structure has successfully coordinated digital transformation efforts across the government, overseeing cybersecurity, artificial intelligence, national technologies and big data, each of which is vital for supporting Turkey’s transition to an effective digital governance model under the presidential system. Historically, e-government initiatives focused on providing individual public services online, but the office expanded this focus to create a comprehensive digital ecosystem.
    \cite{govinsider_2023_turkey}
    
    \item Israel: Ministry of Innovation, Science, and Technology\\
    The Ministry of Science, Technology, and Space promotes projects aimed at encouraging research, focusing on leading strategic research infrastructures. The Ministry is responsible for the development of scientific and technological infrastructure in Israel, advanced research and development, international scientific relations, and the Israel Space Agency. \cite{govil_2023_ministry}
    \cite{govil_2023_ministry}
    
    \item Saudi Arabia: Saudi Data and Artificial Intelligence Authority (SDAIA)\\
    The Saudi Data and Artificial Intelligence Authority is the competent body in the Kingdom responsible for issues related to data and artificial intelligence, including big data. SDAIA is also the national reference for all matters concerning the organization, development, and processing of data and AI. Additionally, its jurisdiction includes issues related to the operation, research, and innovation in the field of data and AI. \cite{sdaia_2023_about}
    \cite{sdaia_2023_about}
    
    \item South Korea: Ministry of Science and ICT\\
    This government institution is responsible for fostering innovation and growth in the country’s science and technology sector, with a particular focus on information and communication technologies (ICT). MSIT plays a crucial role in developing policies that promote research development, support the artificial intelligence industry, and improve digital education, all aimed at strengthening South Korea’s position as a global leader in technology and innovation. \cite{msit_2023_about}
    \cite{msit_2023_about}
    
    \item Japan: Ministry of Internal Affairs and Communications (MIC)\\
    The Ministry of Internal Affairs and Communications regulates the information, communications and postal services sectors, as well as other systems such as administrative organizations, electoral systems and disaster prevention. The Ministry promotes fair competition in these sectors and develops new technological systems in information and communications. The MIC issues licenses to operators and sets tariffs and taxes for regulation. \cite{mic_2023_japan}
    \cite{mic_2023_japan}
    
\end{itemize}

\subsubsection{Authorities Responsible for Developing and Implementing Ethical Standards}
\begin{itemize}
    \item Turkey: Ministry of Industry and Technology\\
    The Ministry of Industry and Technology of Turkey (Sanayi ve Teknoloji Bakanlığı) is a government body responsible for overseeing industry and technology in the country, with the goal of promoting economic growth and innovation. Established in 1984, the ministry plays a crucial role in shaping policies aimed at enhancing competitiveness, sustainable development, and technological progress across various sectors.
    \cite{globaledge_2023_turkey_msit}
    
    \item Saudi Arabia: The Council of Ministers\\
    The Council of Ministers is the highest executive authority in Saudi Arabia, responsible for formulating and implementing the country’s internal and foreign policies across various sectors, including finance, education, and defense. The Council consists of the King as Prime Minister, the Crown Prince as Deputy Prime Minister, and various ministers, with its main offices located in Riyadh and Jeddah. Founded in 1953, it plays a vital role in overseeing state affairs and ensuring the enforcement of laws and regulations through consensus-based decision-making. \cite{saudipedia_2023_council_of_ministers}

\end{itemize}
    
\subsubsection{Authorities Promoting Innovation and the Usage of AI in the Economy}
\begin{itemize}
    \item China: Ministry of Industry and Information Technology (MIIT)\\
    The primary responsibilities of the Ministry include: developing industrial planning, policies, and standards for China; monitoring the daily operations of industrial sectors; promoting the development of key technological equipment and local innovations; overseeing the communications sector; managing the construction of information systems; and protecting China's information security. As an industrial management department, it guides the development of various industries but does not interfere with business operations.
    \cite{globaltimes_2023_ministry_of_industry}
    
    \item Mexico: Secretariat of Economy\\
    The Ministry of Economy is a federal department responsible for economic matters in Mexico, promoting and stimulating the development of a culture that creates better and more transparent business opportunities. The Ministry is also responsible for developing and implementing policies on industrial, foreign, and domestic trade, supply, and pricing in the country. Additionally, it plays an active role in promoting initiatives in sectors such as foreign trade, mining, and commerce.
    \cite{economy_2023_mexico}

    \item Colombia: Ministry of Information Technologies and Communications (MinTIC)\\
    The Ministry of Information Technologies and Communications is responsible for the development, adoption, and promotion of policies, plans, programs, and projects in the information and communication technologies sector. Its functions include expanding and facilitating access for all residents across the national territory.
    \cite{mintic_2023_info}

    \item Poland: Polish Agency for Enterprise Development\\
    The Polish Agency for Enterprise Development (PARP) participates in the implementation of national and international programs funded by EU structural funds, the state budget, and multi-year European Commission programs. As a key organization responsible for creating a favorable business environment in Poland, PARP fosters the formulation and effective implementation of state policies related to entrepreneurship, innovation, and workforce adaptation.
    \cite{parp_2023_info}
    
\end{itemize}

\subsection{Data Regulation}

Data is the foundation for training large language models (LLMs), determining their effectiveness, accuracy, and ability to apply their knowledge in new situations. The regulation of data collection and usage is crucial to ensure transparency and clarity regarding model operations, ethics, and user rights compliance.
Similar to the regulation of model usage, there are key approaches to data transparency. In this section, we will explore these approaches.

\subsubsection{Policy of Extensive Data Availability for AI Models}

This approach involves granting unconditional permission to use large volumes of data to train models, potentially with only minor and comparatively insignificant restrictions.

\begin{itemize}
    \item \textbf{The Case of Japan.}
    The Japanese authorities are actively encouraging unconditional data usage for AI while also establishing national data platforms. This contributes to the rapid growth of local AI technologies, providing Japan with a competitive edge. More importantly, the government has introduced a law allowing AI development companies to use any data for training purposes without concerns about copyright or data sources. The objective is to offer unrestricted access to various data critical for developing effective AI systems. \cite{PrivacyWorld2024, Bunka2024}
    \item \textbf{Example of Saudi Arabia.} 
    As part of its Vision 2030 strategy, Saudi Arabia has initiated efforts to harness open data for AI development, especially in healthcare and energy sectors \cite{gov_sa_2023}. This policy facilitates access to data for researchers, businesses, and startups, allowing them to use, republish, and reuse datasets without facing significant technical or financial barriers, provided they comply with appropriate licensing terms \cite{saudipedia_2024, SDAIA2020_opendata}.
    Although open data initiatives are designed to provide greater access to datasets, there are still legal considerations when using copyrighted materials. Companies must be cautious to avoid complications when integrating third-party data into their AI systems \cite{al_balushi_2023}.
    One noteworthy regulatory point is that intellectual property created without human involvement is considered public domain and exempt from copyright protection. In addition, Saudi Arabia has implemented restrictions on the export of data, permitting it only under specific circumstances—such as fulfilling the Kingdom’s interests or honoring international commitments \cite{al_balushi_2023}.

\end{itemize}

\subsubsection{Partial Data Openness}

This approach allows for a controlled level of data openness, typically for government data, while restricting private or unregulated sources.

\begin{itemize}
    \item \textbf{South Korea case.}
    The South Korean government promotes the use of public data for AI development but forbids the use of uncontrolled private data in models. A National Open Data Portal has been established, offering access to a vast collection of digitized resources (87,000 public datasets and 11,000 open APIs) to encourage innovation in AI. \cite{ying_thian_2021}

    On the other hand, there is a clear prohibition in South Korea on using private or copyright-protected data for training AI models. This is to prevent potential infringements on privacy rights and intellectual property laws. Organizations are encouraged to use publicly available datasets or obtain explicit consent when using private data, ensuring compliance with the Personal Information Protection Act (PIPA). \cite{yulchon_2024, han-ah_yeong-su_2024}

    \item \textbf{The Netherlands case.}
    In the Netherlands, data regulation for artificial intelligence (AI) is a mix of national legislation and European Union directives. The country has introduced clear limitations on using confidential data, such as medical and financial information, to protect individual privacy while facilitating access to less sensitive government data. \cite{uavg_2024}

    Additionally, the Dutch government is actively fostering the development and use of open public data that does not include confidential information. These datasets are available for various purposes, such as enhancing public services and encouraging technological advancements in transportation and urban planning. \cite{NL_EU_2019, autoriteit_persoonsgegevens_2024}
\end{itemize}

\subsubsection{Restriction On Using Data Without Government Permission}

This approach mandates that AI models be trained solely on data approved by the state or specifically generated for specific tasks.

\begin{itemize}
    \item \textbf{Example of Argentina.}
    The government of Argentina implemented strict regulations on AI tools, coupled with limitations on using open data for training, driven by concerns about privacy and model fairness \cite{OECD2021, Datos_Personales_2023}. However, due to the failure of this policy and the slow development of AI in the country \cite{latam_2024}, the new government’s revised strategy aims to go in the opposite direction. The focus will shift towards relaxing regulations and providing broader access to data to facilitate the growth of AI and technology \cite{hamilton_2024}.

\end{itemize}

\subsection{How does the EU view regulations?}
The EU has two major legislative acts addressing AI regulation in this context.
To maintain a balance between innovation and user rights, the European Data Protection Board (EDPB) in its recommendations (GDPR) emphasizes data processing and protection used in models. Another crucial instrument, the proposed Artificial Intelligence Act (AI Act), sets specific requirements for AI systems, including transparency, audits, and compliance with regulations. In this section, we will examine these legislative measures and illustrate how the industry has responded.

\subsubsection {GDPR (General Data Protection Regulation)
)}
\textbf{The European Data Protection Board (EDPB)}, which oversees data protection regulations, has recently published recommendations regarding the use of personal data for training AI models under the General Data Protection Regulation (GDPR). \cite{edpb_2024} In its findings, the EDPB outlines crucial points concerning both the legal bases for data processing and practical steps to mitigate risks. They specifically note that developers may use "legitimate interest" as a justification for processing personal data, but only after conducting a thorough evaluation through a three-step test to assess the necessity of the processing and its impact on user rights. \cite{DataProtectionReport2025}

One of the key aspects is data anonymization. The Board highlights the difficulty of achieving full anonymity, noting that if a model retains identifiable data, it will inevitably be subject to the GDPR. Moreover, organizations intending to use large language models (LLMs) must ensure that the data used for training has been legally collected and processed. Otherwise, the use of these models may be considered unlawful. To minimize risks, the EDPB suggests a series of measures, such as excluding data from vulnerable individuals and refraining from collecting information from sources that explicitly prohibit its use. \cite{edpb_2024}

\subsubsection {EU AI Act}
\label{sec:AI_Act}
\textbf{The AI Act} is a proposed regulation within the European Union aimed at creating a comprehensive legal framework for the regulation of AI technologies, particularly regarding data usage. \cite{EUAIAct2024} One of the key features of the Act is the classification of AI systems based on their risk levels. High-risk systems (used in areas like critical infrastructure management, education, recruitment assessments, public services access, justice, and jurisprudence) — which could include large language models (LLMs) — are subject to more stringent data usage requirements. These include detailed rules for handling personal data during model training. \cite{lomas_2024}

Transparency is another crucial principle. Developers of high-risk systems must ensure that users receive clear and transparent information about how their models utilize personal data. This approach strengthens accountability and builds trust in the technology. \cite{FutureOfLifeInstitute2024}
Moreover, the law mandates the establishment of compliance mechanisms. Organizations must regularly conduct audits and data processing impact assessments to confirm their practices align with the AI Act's principles. These steps are intended to foster a safer and more ethical environment for the development and use of AI systems. \cite{EUAIAct2024}

\subsubsection {Industry Reaction}

The AI laws and recommendations adopted by the European Union have triggered a diverse range of responses from major technology companies, especially from the United States. Some see this legislation, designed to establish a regulatory foundation for AI, as an essential move toward the responsible development of technology. In contrast, others worry that it could hinder innovation and competitiveness.

\begin{itemize}

    \item \textbf{Innovation slowdown concerns}
Executives at companies like Amazon and Meta have voiced concerns that the EU AI Act may stifle technological progress. Werner Vogels, Amazon's Chief Technology Officer, highlights the importance of a balanced regulatory approach, warning that excessive restrictions could drive innovation out of Europe. He compares the potential effects of the AI Act to those of the General Data Protection Regulation (GDPR), which he believes is excessively complicated and burdensome for businesses \cite{euronews2023potentially}.

Yann LeCun, Head of AI at Meta, questions whether strict AI regulations are appropriate. He asserts that current AI technologies are still far from surpassing human intelligence and advocates for prioritizing the safety of future advancements instead of implementing strict rules right now \cite{knight2024meta}.

In an open letter to EU legislators, over 160 executives from major companies such as Siemens, Carrefour, Renault, and Airbus cautioned that the AI Act could lead to significant compliance costs and disproportionate liability risks. They claim that these burdens might cause innovative companies and investors to avoid Europe, creating a productivity gap compared to regions like the United States \cite{kang2023eu}.

\item \textbf{Mixed feelings among European companies}

European tech companies have mixed reactions to the AI Act. Some leaders support its risk-based framework, which sorts AI applications by their potential risks. However, there are concerns about the strict demands for high-risk uses, which might unduly impact small businesses and startups. France Digitale, for instance, has warned that the Act’s stringent measures could place substantial financial burdens on emerging companies. They advocate for a more nuanced regulatory approach focusing on specific applications rather than the technology as a whole \cite{matthews2023ai, euronews2023potentially}.

\item \textbf{Broader Consequences}

The enactment of the AI Act is a crucial development in regulating artificial intelligence, being the first comprehensive legal framework dedicated to AI governance. However, many leaders in the tech industry are concerned about the potential economic effects of this legislation. For instance, the Computer and Communications Industry Association has raised alarms that the rigid regulations could stifle competition and innovation in Europe, as well as decrease investments in local AI ventures \cite{matthews2023ai, barry2023why}.

Several industry executives have signed open letters expressing concerns that such regulations might undermine Europe’s technological sovereignty, leading companies to relocate their operations outside the EU \cite{matthews2023ai, euronews2024eu}.

\end{itemize}

The EU's AI Act is an innovative attempt at regulating artificial intelligence, sparking debates about its impact on innovation and competitiveness. While some view it as a necessary tool to ensure the responsible development of AI, others fear that the law will create barriers that could hinder growth and lead to a brain drain from Europe. The full impact of this legislation on Europe and the global tech industry will only unfold in time.

\section{International Cooperation}

Large language models have become a catalyst for international cooperation in many regions. In Europe and other parts of the world, programs have been established to provide grants for developing such technologies. Moreover, tech giants have intensified collaboration with companies and governments to create national and international AI projects. In this section, we will discuss these trends and provide several examples from different regions.

\subsection{EU Programs} \label{subs:intr_programs_eu}

In this section, we will discuss programs in Europe aimed at the development of LLMs and Ukraine's potential participation.

\label{sec:eu_programs}
\begin{itemize}

    \item \textbf{CLARIN.} CLARIN is a European Research Infrastructure Consortium (ERIC) established in 2012. Its primary function is to provide a distributed digital infrastructure with centers located across Europe and throughout the world. The goal of CLARIN is to collect as much linguistic data in digital format as possible and to develop tools for language processing \cite{a2020_clarin}. This organization makes a significant contribution to the development of LLMs.

    Any country can become a full member of CLARIN by first obtaining observer status, provided that the share of non-EU member countries in the organization does not exceed 50\% \cite{wynne_2020_what, a2020_national}.

    CLARIN has extensive experience collaborating with non-EU countries, as illustrated by Iceland, Norway, and South Africa, full members of CLARIN \cite{a2020_national}. Additionally, CLARIN has assisted in developing language models for Macedonian and Serbian, even though these countries are not EU members or even associated with CLARIN \cite{a2021_language, a2020_national}.

    CLARIN is already compiling a list of initiatives that support the research community in Ukraine \cite{a2022_initiatives}, and it has a research center in Germany specializing in natural language processing (NLP) for the Ukrainian language \cite{a2025_ukrnlpcorpora}. Moreover, after the onset of the full-scale invasion, CLARIN expressed its support for Ukraine by ceasing operations of its centers in Belarus \cite{a2022_clarin}. This demonstrates the potential for partnership with Ukraine within CLARIN and the possibility of establishing relevant centers to support and develop technologies for language processing in Ukraine.

    Currently, the organization has 24 full members and 2 members with an observer status \cite{a2020_national}. Full members can participate in funded projects, receive technical support, and contribute to forming CLARIN’s policies and strategies. Observers must pay an annual fee and receive support from their national funding institutions \cite{wynne_2020_what}.

    CLARIN develops various language models and does not focus exclusively on generative AI. Their developments include models for morphological analysis, machine translation, named entity recognition (NER), and other applications \cite{a2021_language}.

    Collaboration with CLARIN yields tangible results. For instance, the Polish Office of Competition and Consumer Protection (Urząd Ochrony Konkurencji i Konsumentów) sought to leverage CLARIN’s capabilities to develop a system automatically detecting abusive clauses in legal agreements. The goal of this system is to protect consumers from corporate abuse. As a result, they jointly developed a system that identifies suspicious documents with an F-score of 87\%, which is said to be a very high level of accuracy \cite{netguru_2022_n19}.

    \item \textbf{Horizon Europe.} Horizon Europe is the EU's main program for funding research and innovation, which includes artificial intelligence. According to the Multiannual Financial Framework Midterm Review (MTR), the estimated funding for Horizon Europe from 2021 to 2027 is €93.5 billion \cite{europeancommission_2021_horizon}.

    Under the AI Innovation Package, Horizon Europe and Digital Europe will jointly invest €1 billion each year from 2024 to 2027 to support the development of generative AI technologies, including generative language models \cite{europeancommission_2024_commission}. Overall, the European Commission aims to reach €20 billion in AI investments by the end of this decade \cite{europeancommission_2024_europeanresearch}. For instance, just in April 2024, several calls for proposals were announced, allocating €50 million for the development of large AI models and €15 million for improving AI transparency and reliability \cite{europeancommission_2024_europeanresearch}.

    The Horizon Europe program is open to all legal entities registered in an EU Member State or a country associated with Horizon Europe. These organizations can participate in the program and receive funding \cite{danishagencyforhighereducationandscience_2024_who}. Ukraine already has an association agreement with Horizon Europe. As of December 20, 2023, the National Research Foundation of Ukraine signed a grant agreement with the European Commission, resulting in the opening of the first Horizon Europe office in Ukraine \cite{horizon}.

    Horizon Europe invests significant resources in the development of generative language models. For example, on January 5, 2024, the Large AI Grand Challenge was launched, inviting participants to develop a generative LLM \cite{eugrantsandfunding_2024_call}. As a result of this challenge, teams from France, Portugal, Latvia, and Belgium were awarded €1 million each in funding, along with 8 million GPU hours on EuroHPC JU supercomputers, to train their generative language models over the following year \cite{europeanhighperformancecomputingjointundertaking_2024_winners}.

    Additionally, on December 2, 2024, the EuroLLM-9B LLM was released \cite{eurollmteam_2024_eurollm9b}. This model was developed through collaboration among private companies, institutes, universities, and a private research lab, which were based in Portugal, the UK, France, the Netherlands, and South Korea. The development was supported by EuroHPC JU and funded as part of the UTTER project under Horizon Europe, with a total budget exceeding €4 million \cite{unifiedtranscriptionandtranslationforextendedreality_2022_transforming}. EuroLLM-9B supports all 24 official European languages and 11 major international languages, with a larger and higher-quality model currently under development \cite{eurollmteam_2024_eurollm9b}.

    Among other projects launched under UTTER is a Croatian LLM \cite{a2024_croatian}. This project's goal is to collect data and train the model, with plans to publish it in the repository of the Croatian branch of CLARIN \cite{a2024_hrclarin}. This demonstrates the project's commitment to developing sovereign LLMs for individual countries.

    \item \textbf{Digital Europe.} The Digital Europe program is an EU funding initiative aimed at aiding the deployment of digital technologies for businesses, citizens, and government agencies \cite{europeancommission_2021_digital}. Unlike Horizon Europe, which focuses on innovation and development, Digital Europe emphasizes implementation and real-life use. Currently, the program's total budget has increased to over €8.178 billion \cite{europeancommission_digital}, with approximately €2.1 billion allocated for AI \cite{europeancommission_2021_digital}.

    Ukraine already has a limited association with the Digital Europe program \cite{a2024_list}, signed in September 2022. This association is limited to objectives 1, 2, 4, and 5, which include high-performance computing (objective 1) and artificial intelligence (objective 2) \cite{a2018_legal}. This allows Ukraine to participate in relevant programs.

    The Digital Europe program allocated €1.9 billion to EuroHPC JU \cite{europeancommission_2023_the}, enabling the creation of supercomputers for AI training. Among these is the MareNostrum 5 supercomputer in Spain, used to train the EuroLLM-9B language model mentioned above \cite{eurollmteam_2024_eurollm9b}. Additionally, the program funds the Alliance for Language Technologies (ALT-EDIC), coordinated by France, which focuses on collecting, training, and distributing language processing technologies, including generative language models \cite{europeancommission_2025_eu}. During the formation of this alliance, Ukraine was among the countries whose organizations could participate \cite{euroaccess_2024_alliance}, demonstrating readiness for collaboration with Ukrainian institutions.

    \item \textbf{EuroHPC JU.} The European High Performance Computing Joint Undertaking (EuroHPC JU) is a collaborative initiative between the EU, European countries, and private partners aimed at developing a pan-European supercomputing infrastructure and supporting research and innovation activities \cite{europeancommission_2023_the}.

    In the regulation adopted by the European Council in July 2021, €7 billion was allocated to this program for the 2021–2027 period \cite{counciloftheeuropeanunion_2021_european}. This funding includes contributions from Digital Europe (€1.9 billion), Horizon Europe (€900 million), the Connecting Europe Facility (€200 million), private partners (€900 million), and equal contributions from member states.

    Supercomputers built under the EuroHPC initiative frequently train artificial intelligence models, particularly LLMs, which require significant computational resources. Projects such as EuroLLM-9B (MareNostrum 5, Spain) \cite{eurollmteam_2024_eurollm9b}, TrustLLM (MareNostrum 5, Spain) \cite{eurohpcjointundertaking_2023_trustllm}, the Large AI Grand Challenge (LUMI, Finland, and Leonardo, Italy) \cite{aiboost_2023_large}, Latxa (Leonardo, Italy) \cite{basquecenterforlanguagetechnology_2024_latxa}, and others involved in developing LLMs have been made possible through infrastructure provided by EuroHPC JU.

    Membership in EuroHPC is not restricted to EU countries. For instance, Norway, North Macedonia, Serbia, Iceland, and others participate in this initiative \cite{europeancommission_2023_the}. Specifically, researchers from academic institutions, research institutes, public authorities, and entities located in an EU member state, a EuroHPC participant state, or a country associated with the Digital Europe or Horizon Europe programs can access EuroHPC supercomputers acquired after 2020 \cite{eurohpcjointundertaking_2023_access}.

    Only a few countries have the resources to independently develop supercomputers, making EuroHPC a critical source of support. A prominent example is AI Factories, which will be further discussed below. In the future, Ukraine could participate in similar initiatives by initiating large-scale infrastructure projects within its territory.

    \item \textbf{AI Factories.} AI Factories leverage the supercomputing resources of EuroHPC for the development of generative AI models \cite{europeancommission_2024_ai}. The European Commission prioritized the establishment of such AI Factories in a policy package adopted on January 24, 2024 \cite{europeancommission_2024_commission}. By December 2024, EuroHPC had selected seven candidate countries for creating the first AI Factories. The construction of these supercomputers will involve contributions from 15 EU member states, and 2 EuroHPC participant states. This includes Norway and Turkey, non-EU member states that will support the development of supercomputers in Finland and Spain \cite{europeancommission_2024_ai}.

    Some countries also plan to build their AI Factories independently. For example, the Pharos infrastructure will be added to the Daedalus supercomputer in Greece for €30 million, with €15 million provided by EuroHPC \cite{greeknewsagenda_2024_greeces}. This supercomputer will be used to create a language model that will serve as the next iteration of the Meltemi model, released in 2024. Investments in infrastructure are key to developing sovereign language models, making this initiative a potential example of Ukraine's future strategy.

\end{itemize}

\subsection{International Cooperation in Other Regions}

In this section, we will examine additional examples of international cooperation in regions outside the EU. Through these examples, we aim to illustrate potential approaches to regional collaboration that are not necessarily conducted within the framework of the European Union.

\begin{itemize}

    \item \textbf{Global Partnership on Artificial Intelligence.} The Global Partnership on Artificial Intelligence (GPAI) is a multilateral initiative aimed at supporting advanced research and applied activities in priority AI areas \cite{theglobalpartnershiponartificialintelligence_2020_about}.

    Launched in June 2020, it comprises 44 countries \cite{organisationforeconomiccooperationanddevelopment_global} and aims to implement the recommendations of the Organisation for Economic Co-operation and Development (OECD) \cite{theglobalpartnershiponartificialintelligence_2020_about}. During its initial years, GPAI experts will collaborate in four working groups: responsible AI, data governance, the future of work, and innovation and commercialization.

    Membership in GPAI is open to countries, including those with transitioning or developing economies. Given that Ukraine has shown intentions to join the OECD in recent years \cite{organisationforeconomiccooperationanddevelopment_ukraine}, it could also consider participating in this initiative to advance its AI capabilities.

    \item \textbf{BigScience Workshop.} The BigScience Workshop was founded in early 2021. Its primary founders include Thomas Wolf, CEO of Hugging Face, as well as Stéphane Requena (Grand Équipement National de Calcul Intensif, GENCI) and Pierre-François Lavallée (Institut du Développement et des Ressources en Informatique Scientifique, IDRIS), both associated with the French supercomputer Jean Zay \cite{bigscienceworkshop_short}.

    The BigScience project is inspired by collaborative scientific models such as CERN and the Large Hadron Collider (LHC), where open scientific collaboration facilitates the creation of large-scale artifacts beneficial to the entire research community \cite{a2021_bigscience}. Through this project, language models such as CamemBERT \cite{martin_2020_camembert} and BLOOM \cite{bigscienceworkshop_bloom} have already been created. The former is designed for the French language, while the latter supports all major languages and serves as a competitor to ChatGPT.

    The BigScience Workshop is viewed as a platform for conducting international collaborative research. In addition to creating and sharing research artifacts, the project aims to consolidate knowledge that will enable similar experiments in the future \cite{bigscienceworkshop_introduction}.

    \item \textbf{Singapore and Southeast Asia.} Singapore has become a regional hub for the development of LLMs. Project SEALD \cite{aisingapore_2024_southeast}, which collects data for the Singaporean language model SEA-LION \cite{aisingapore_2023_sealion}, is a joint initiative of AI Singapore, VISTEC from Thailand, and XFORM from Indonesia. The SEA-LION language model, developed by AI Singapore and published in open source, supports all major languages spoken in Southeast Asia \cite{aisingapore_2023_sealion}.

    \item \textbf{France and Germany.} France has expressed its ambitions to become a leader in artificial intelligence in Europe, aiming to counterbalance American language models. To achieve this, it actively promotes collaboration with other European countries. Notably, the French research institute Inria and the German research institute DFKI share a partnership and are engaged in five joint projects \cite{inria_2020_the}.

    \item \textbf{AmericasNLP Workshop.} The AmericasNLP Workshop \cite{americasnlp_2025_thrid} annually brings together researchers from various Latin American countries to collaborate on language processing projects focused on the languages of Indigenous peoples of the Americas. Their efforts go beyond the technical aspects, incorporating interdisciplinary collaboration that unites linguists, computer scientists, and community representatives to address the challenges of low-resource languages.

    \item \textbf{UlizaLlama.} In Kenya, the UlizaLlama (AskLlama) \cite{jacarandahealth_2024_jacaranda} project has been launched to provide free medical consultation for people without direct access to healthcare. A key feature of this chatbot is its advanced understanding of African languages, which are underrepresented in popular language models. Jacaranda Health \cite{jacarandahealth_2023_about}, the organization developing this model, collaborates with government institutions such as the Ministry of Health in Kenya, the Ministry of Health in Eswatini, and the Ghana Health Service, among others. The project is also supported by numerous international funds, including USAID, IDRC/CRDI, CRI Foundation, and others.

\end{itemize}

\subsection{Collaboration with Major Companies \& Corporations}
\label{sec:BigCompanies}

In this section, we will showcase how international corporations and companies have partnered with countries to create language models tailored to the specific languages and challenges of these nations. The purpose of this section is to demonstrate the success of such projects, highlight the ability of private companies to develop artificial intelligence solutions for specific countries and showcase the unique aspects of such collaborations.

\begin{itemize}

    \item \textbf{NVIDIA and Spain.} The Barcelona Supercomputing Center (Centro Nacional de Supercomputación, BSC-CNS) and NVIDIA, a leading corporation specializing in developing computing technologies for AI, have signed a collaboration agreement aimed at jointly developing innovative solutions over the next five years, with a significant focus on artificial intelligence. Among the initial projects for this partnership are LLMs \cite{barcelonasupercomputingcenter_2023_collaboration}.

    The MareNostrum 5 supercomputer, which is part of the BSC-CNS and is equipped with approximately 4,500 NVIDIA H100 GPUs, which are among the most advanced in the AI industry \cite{nvidiadeveloper_2024_nvidia}. The EuroLLM-9B language model was trained on this supercomputer \cite{eurollmteam_2024_eurollm9b}, as previously mentioned in subsection \ref{subs:intr_programs_eu}.

    \item \textbf{NVIDIA and India.} In India, NVIDIA developed the Nemotron-4-Mini-Hindi-4B language model, specifically trained for Hindi \cite{dhupar_2024_india}. According to NVIDIA, this specialized model leads in multiple accuracy tests among AI models with up to 8 billion parameters. It is a foundation for companies aiming to create their own language models in India for various applications.

    One of the first clients to utilize this service was Tech Mahindra, one of the largest IT and consulting companies, which launched the Indus 2.0 project. In the first phase of this project, they plan to develop a language model for Hindi and 37 dialects, with subsequent phases focusing on additional languages and even more dialects \cite{a2024_the}. As a result of this initiative, the model was released to open source.

    Additionally, Tech Mahindra announced the establishment of an AI Center of Excellence based on platforms and applications developed by NVIDIA. When discussing the goal of this collaboration, they emphasized the priority of advancing sovereign large language models \cite{techmahindra_2024_tech}.

    Beyond the partnership with Tech Mahindra, NVIDIA’s services were also utilized by Sarvam AI, an Indian company specializing in generative AI, to build the Sarvam-1 language model \cite{sarvamai_2024_sarvam, dhupar_2024_india}. This model supports 10 major Indian languages as well as English and outperforms language models such as Google’s Gemma-2-2B \cite{googleaifordevelopers_gemma} and Meta’s Llama-3.2-3B \cite{meta_llama} in several benchmarks.

    \item \textbf{Google and Singapore.} A study by Access Partnership, commissioned by Google, highlights yet another example of successful collaboration between the public and private sectors \cite{accesspartnership_2024_strengthening}. This example is the joint initiative AI Trailblazers, launched in collaboration with the Ministry of Communications and Information (MCI), public initiatives Digital Industry Singapore (DISG), Smart Nation and Digital Government Office (SNDGO), and Google Cloud. The goal of the AI Trailblazers project is to bring together marketers, technologists, entrepreneurs, and venture capitalists to accelerate the adoption of generative AI in Singapore. Participants in the program, equipped with the necessary resources, implemented generative AI solutions to address specific tasks in their workplace environments \cite{singaporeeconomicdevelopmentboard_2023_mci}.

    Specifically, as part of this project, Nanyang Polytechnic developed Course AutoBot, a tool powered by LLMs, to assist teachers in course creation. By accelerating the course development process of educators, the tool leaves them significantly more time to interact with students and provide a more valuable learning experience \cite{accesspartnership_2024_strengthening}.

    \item \textbf{NVIDIA and Indonesia.} Another initiative developed by NVIDIA is the Sahabat-AI collection of language models \cite{sahabatai_2017_sahabatai, gomes_2024_indonesia}, designed specifically for the Indonesian market and built in collaboration with the Indian company Tech Mahindra.

    Sahabat-AI is a collection of Indonesian LLMs made available as open-source tools, enabling local companies, government agencies, universities, and research centers to create their own generative AI-based applications.

    The Hippocratic AI company \cite{hippocraticai_foundation} was among the first to use Sahabat-AI for medical applications. They developed digital agents capable of performing nursing tasks and providing necessary medical advice.

    \item \textbf{Saudi Arabia.} Saudi Arabia collaborates with international companies such as IBM, Huawei, AWS, and others. These partnerships focus on developing LLMs and the infrastructure to support them.

    For instance, through its collaboration with IBM, the Saudi Data and Artificial Intelligence Authority (SDAIA) launched an Arabic large language model designed for use by businesses and government agencies for text-related tasks \cite{ibmnewsroom_2024_through}.

    Additionally, an agreement between SDAIA and Amazon Web Services (AWS) will bring over \$5.3 billion in investments to build an infrastructure capable of training AI models \cite{amazonpresscenter_2024_aws}.

    Finally, the support from Huawei Cloud, a product of the Chinese company Huawei, enabled Saudi Arabia to develop a language model that will power more than 20 applications based on its framework \cite{kudakwashemuzoriwa_2024_huawei}.

    \item \textbf{Microsoft and Mexico.} The Monterrey Institute of Technology and Higher Education (Tec de Monterrey) partnered with Microsoft to create TECgpt, an AI-powered platform to transform education. The goal of this project is to personalize education to meet students' needs, enhance the learning process, stimulate teachers' creativity, and save time on repetitive tasks \cite{microsoft_2024_tecnolgico}. This language model is based on GPT-4o by OpenAI, a company into which Microsoft has reportedly invested around \$13 billion \cite{published_2023_openai}.

    Furthermore, Microsoft announced an additional \$1.3 billion investment in cloud infrastructure and AI in Mexico, with the aim of driving inclusive economic growth through technology and educational programs \cite{a2024_microsoft}.

    \item \textbf{NVIDIA and Israel.} In 2023, NVIDIA announced the construction of Israel-1, an AI supercomputer estimated to cost hundreds of millions of dollars \cite{datacenterdynamics_2023_nvidias}. Completed in 2024, this supercomputer became the most powerful in the country, ranking 34th globally \cite{newtechmagazinesgroupltd_2024_nvidias}. Currently housed in NVIDIA’s Israeli data center, the supercomputer is exclusively available by NVIDIA’s research and development teams and selected unnamed partners.

\end{itemize}

\section{Conclusions}

In this study, we have provided multiple examples illustrating the involvement of countries worldwide in developing their own LLMs. We focused on relevant applications, including national security and funding strategies, mainly through EU programs with which Ukraine is associated. Additionally, we examined the risks associated with AI and highlighted examples of regulations to mitigate them.

The examples discussed in this study provided an overall view of the opportunities and challenges associated with LLMs and AI in general. Specifically, this research broadened the focus from the key players in AI development. This approach allowed us to analyze the experiences of countries whose paths and strategies may be more relevant to those of Ukraine than of nations holding leading global roles. We hope this study will serve as a helpful resource in shaping the course of action Ukraine chooses for the future development of large language models and AI in general.

\section{Acknowledgments}
The authors are sincerely grateful to OpenBabylon organization for the financial support in creating this report and to Oleksii Molchanovsky and Mariana Romanyshyn for their detailed feedback and comments.
\newpage

\bibliographystyle{abbrv}
\bibliography{sample}

\end{document}